\documentclass[12pt,a4paper]{article}
\usepackage{graphics,epsfig}
\begin{document}
\newcommand{\goo}{\,\raisebox{-.5ex}{$\stackrel{>}{\scriptstyle\sim}$}\,}
\newcommand{\loo}{\,\raisebox{-.5ex}{$\stackrel{<}{\scriptstyle\sim}$}\,}

\begin{center}
{\Large \bf Statistical approach for supernova matter.}
\end{center}
\vspace{0.5cm}
\begin{center}
{\Large A.S.~Botvina$^{a,b}$ and I.N.~Mishustin$^{b,c}$}\\
\end{center}
\begin{center}
{\it
 $^a$Institute for Nuclear Research, Russian Academy of Sciences, 
117312 Moscow,
 Russia\\
 $^b$Frankfurt Institute for Advanced Studies, J.W. Goethe University, D-60438
     Frankfurt am Main, Germany\\
 $^c$Kurchatov Institute, Russian Research Center, 123182 Moscow, Russia\\
}
\end{center}
\normalsize

\vspace{0.3cm}
\begin{abstract}
We formulate a statistical model for description of nuclear composition 
and equation of state of stellar matter at subnuclear densities and 
temperature up to 20 MeV, which are expected during the collapse and 
explosion of massive stars. The model includes 
nuclear, electromagnetic and weak interactions between all kinds of 
particles, under condition of statistical equilibrium. 
We emphasize importance of realistic description of the 
nuclear composition for understanding stellar dynamics and nucleosynthesis. 
It is demonstrated that the experience accumulated in 
studies of nuclear multifragmentation reactions can be used for better 
modelling properties of stellar medium. 
\end{abstract}

\vspace{0.2cm}


\vspace{0.5cm}

{\bf 1. Introduction}\\

In violent nuclear reactions 
strong interaction between many nucleons leads to a fast equilibration. 
This short-range interaction is responsible for a sharp 
freeze-out when the inter-particle distance becomes larger than 
the interaction range. For these reasons statistical models have proved 
to be very successful for interpretation of nuclear reactions at various 
energies. They are widely used for description of fragment production 
when one or several equilibrated sources can be identified. 
Originally this concept was proposed for compound nucleus decays, such 
as evaporation or fission of excited nuclei \cite{Bohr}. Recently, it 
was demonstrated that the concept of equilibrated source can even be 
effectively used 
for more violent multifragmentation reactions leading to production of 
many fragments \cite{SMM}. On other side, the statistical equilibrium 
is expected in many astrophysical processes, when the characteristic 
time for formation of nuclei is much shorter than the time-scale of 
these processes. For example, one of the most spectacular astrophysical 
events is a type II supernova explosion, with a huge energy release of 
about several tens of MeV per nucleon \cite{Brown,Bethe}. 
When the core of a massive star collapses, it reaches densities several
times larger than the normal nuclear density $\rho_0=0.15$ fm$^{-3}$.
The repulsive nucleon-nucleon interaction gives rise to a bounce-off
and creation of a shock wave propagating through the in-falling stellar
material. This shock wave is responsible for the ejection of a star 
envelope that is observed as a supernova explosion. During the collapse 
and subsequent explosion the temperatures $T\approx (0.5\div 10)$ MeV 
and densities $\rho \approx (10^{-6}\div 2) \rho_0$ can be reached. It 
is widely believed that the nuclear statistical equilibrium should be 
reached under these conditions. As shown by many theoretical studies, 
a liquid-gas phase transition should take place in nuclear matter under 
such conditions. 

As discussed by several authors (see e.g. 
\cite{Janka,Thielemann,Sumi,Burrows}), 
there are problems in producing successful explosions in hydrodynamical 
simulations of the core-collapse supernovae, even when neutrino heating and 
convection effects are included. The hope is that full 3-$d$ simulations 
will help to solve this problem \cite{3d}. 
On the other hand, it is known that nuclear composition is extremely important
for understanding the physics of supernova explosions. In particular, the
weak reaction rates and energy spectra of emitted neutrinos are very sensitive
to the presence of heavy nuclei (see e.g. \cite{Ring,Hix,Langanke,Horowitz}).
This is also true for the equation of state (EOS) used in hydrodynamical
simulations since the shock strength is diminished by dissociation of heavy 
nuclei. 

Nuclear reactions in a supernova environment are especially important because 
supernova explosions may be considered as breeders for creating 
chemical elements. According to present understanding, there are three 
main sources of chemical element production in the Universe. 
The lightest elements (up to He, and, partly, Li) are formed during the 
first moments of the Universe expansion, immediately after the Big Bang. 
Light elements up to $^{16}$O can be produced in thermonuclear reactions 
in ordinary stars like our Sun, while 
heavier elements up to Fe and Ni can be formed in heavy stars at the 
end of the nuclear burning epoch. It is most likely that heavy 
elements up to U were synthesized in the course of supernova explosions. 
Pronounced peaks in the element abundances can be explained by neutron 
capture reactions in s- and r-processes \cite{Cowan,Qian}. It is believed 
that suitable conditions for the r-process were provided by free neutrons 
abundantly produced in supernova environments together with appropriate 
seed nuclei. 

The EOS of supernova matter is under investigation for more than 25 years. 
One of the first EOS, frequently used in supernova simulations, was obtained 
in refs. \cite{Lamb,Lattimer} many years ago. It includes both light 
and heavy nuclei in statistical equilibrium. 
However, it does not include the whole ensemble of hot heavy nuclei, 
replacing them by a single  ``average'' nucleus. The same assumption 
within a relativistic mean-field approach was used in the EOS of ref. 
\cite{Shen}. As was already pointed out by many authors 
\cite{Japan,Botvina04,Langanke_structure,Janka2} 
this assumption is not sufficient for accurate treatment of the 
supernova processes. We think that this kind of approximation 
can distort the true statistical ensemble in many cases. 
There are other statistical calculations which consider the ensemble 
with different nuclear species, but in the partition sum they include 
only nuclei in long-lived states known from 
terrestrial experiments (see refs.~\cite{Japan,Langanke_structure,Janka2}). 
Also, for description of unknown neutron-rich hot nuclei only properties 
(e.g., the symmetry energies) of cold and slightly excited isolated nuclei 
have been used up to now. These assumptions are not justified for 
supernova environments characterized by relatively high temperatures 
(up to 10 MeV) and densities of electrons and baryons in the range 
$\rho \approx 10^{-4}\div10^{-1} \rho_0$. We believe, in order to achieve 
a more realistic description of supernova matter, it is necessary to use rich 
experience accumulated in recent years by the nuclear community in studying 
highly excited equilibrated systems in nuclear reactions. In particular, 
multifragmentation reactions provide valuable information about hot nuclei 
in dense surrounding of nucleons and other nuclei, which in many aspects is 
similar to supernova interior \cite{Botvina05}.

\vspace{0.5cm}

{\bf 2. Nuclear reactions in supernova environments.}\\

{\bf 2.1 Studying equilibrated nuclear matter in laboratory. } \\

Properties of strongly-interacting nuclear matter are studied experimentally 
and theoretically for a long time. At present there exist consensus 
about phase diagram of nuclear and neutron matter (see, e.g., refs. 
\cite{Mosel76,Lamb78,Lamb}). It is shown schematically in Fig.~1 for 
symmetric nuclear matter, for range of 
densities and temperatures expected in the Supernova II explosions. It is 
commonly accepted that this diagram contains a liquid-gas phase transition. 
From this phase diagram one can conclude that nuclear 
matter at densities $\rho \approx 0.3-0.8\rho_0$ and temperatures $T < T_c$ 
should be in the mixed phase. This phase is strongly inhomogeneous with 
intermittent dense and dilute regions. In the case of electrically-neutral 
matter this mixed phase may have different topologies 
such as spherical droplets, cylindrical nuclei, slab-like configurations 
and others. These configurations are generally referred 
to as nuclear 'pasta' phases \cite{Pethick}, which were recently under 
intensive theoretical investigation \cite{Pethick2,Horowitz2,Watanabe}. 
However, in the coexistence region at lower densities, $\rho < 0.3\rho_0$, 
which are considered in this paper, the nuclear matter breaks up into 
compact nuclear droplets surrounded by nucleons. These relatively low 
densities dominate during the main stages of stellar collapse and explosion. 
Description of nuclear composition in this region requires theoretical 
extrapolation of nuclear properties to these extreme conditions. 
As became obvious after intensive experimental studies of nuclear 
multifragmentation reactions, they proceed through 
formation of thermalized nuclear systems characterized by subnuclear 
densities $\rho \sim 0.1 \rho_0$ and temperatures of 3--8 MeV. Thermodynamic 
conditions associated with these reactions are indicated 
by the shaded area in Fig.~1. This gives us a chance to extract properties 
of hot nuclei in the environment of other nuclear species directly from 
the experimental multifragmentation data and then use this information 
for more realistic calculations of nuclear composition of stellar matter. 
We have also shown isentropic trajectories with the entropy per baryon 
(S/B) of 1, 2, and 4 units (see section {\bf 5.2}) typical for supernova 
explosions. One can see, for example, that an adiabatic collapse (and 
expansion) of stellar matter with typical entropies of 1-2 units per baryon 
passes exactly through the multifragmentation domain. 


Nuclear multifragmentation, i.e. break-up of hot heavy nuclei into many 
fragments, was under intensive investigation during 
the last 20 years. It was solidly established by both theoretical and 
experimental studies that this channel dominates at high excitation energies, 
above 3--4 MeV per nucleon, replacing sequential evaporation 
and fission of the compound nucleus, which is conventional mechanism 
at low excitation energies. In this respect, multifragmentation is 
a universal process expected in all types of nuclear reactions, induced by 
hadrons, heavy ions, and electromagnetic collisions, where the nucleus 
receives a high excitation energy. The Statistical Multifragmentation Model 
(SMM) \cite{SMM} is one of the most successful models used for theoretical 
interpretation of these reactions. Some examples, how low-density 
equilibrated nuclear systems can be produced, and how well their decay can 
be described within this statistical approach, can be found in refs. 
\cite{SMM,Botvina90,ALADIN,EOS,MSU,INDRA,FASA,Dag}. Recently, experimental 
evidences have appeared \cite{LeFevre,Iglio,Souliotis} that the symmetry 
energy of hot fragments produced in multifragmentation reactions is 
significantly 
reduced as compared with the value in cold nuclei. The surface and bulk 
energy of nuclei in this hot and dense environment can be  modified 
too \cite{Botvina06,bulk}. These conclusions, made after appropriate 
analyses of experimental data, suggest that the same modifications will 
occur in stellar matter at similar densities and temperatures. Actually, 
the "in-medium" modifications are quite expected, since the nuclei will 
interact with the surrounding matter, and, therefore, should change 
their properties. 
These modifications may have important consequences for nuclear composition, 
equation of state, and weak reaction rates on these nuclei.

\vspace{0.5cm}

{\bf 2.2 Nuclear and electro-weak reaction rates.} \\

In the supernova environment, as compared to the 
nuclear reactions, several new important ingredients should be taken 
into consideration. First, the matter at stellar scales must be 
electrically neutral and therefore electrons should be included to 
balance a positive nuclear charge. 
Second, energetic photons present in hot matter may change 
nuclear composition via photo-nuclear reactions. 
And third, the matter is irradiated by a strong neutrino wind from the 
protoneutron star.

We consider macroscopic volumes of matter consisting of various nuclear 
species with mass number $A$ and charge $Z$, 
$(A,Z)$, nucleons $(n=(1,0)$ and $p=(1,1))$, electrons $(e^-)$ 
and  positrons $(e^+)$ under condition of electric neutrality. 
There exist several reaction types responsible for the chemical composition 
in supernova matter. At low densities and 
temperatures around a few MeV the most important ones are: 
1) neutron capture and photodisintegration of nuclei, which proceed via 
production of a hot compound nucleus 
\begin{eqnarray}
(A,Z)+n\rightarrow (A+1,Z)^{*}\rightarrow (A+1,Z)+\gamma~,~...\nonumber\\
(A,Z)+\gamma\rightarrow (A,Z)^{*}\rightarrow (A-1, Z)+n~,~...
\end{eqnarray}
2) neutron and light charged particle emission (evaporation) by the hot nuclei
\begin{equation}
(A,Z)^{*}\rightarrow(A-1,Z)+n~,~ (A,Z)^{*}\rightarrow(A-1,Z-1)+p~,~...
\end{equation}
and 3) weak processes
induced by electrons/positrons and
neutrinos/antineutrinos
\begin{equation} \label{enu}
(A,Z)+e^-\leftrightarrow (A,Z-1)+\nu~,
~(A,Z)+e^+\leftrightarrow (A,Z+1)+\tilde{\nu}~,
\end{equation}
which transfer protons to neutrons and vice versa. 
There are many other reactions not shown here, which are naturally taken 
into account within the assumption of statistical equilibrium. 
The characteristic reaction times for neutron capture,
photodisintegration of nuclei and nucleon emission are defined as
\begin{eqnarray} \label{rate}
\tau_{\rm cap}=
\left[\langle\sigma_{nA}v_{nA}\rangle \rho_n\right]^{-1}~,\nonumber\\
\tau_{\gamma A}=
\left[\langle\sigma_{\gamma A}v_{\gamma A}\rangle \rho_\gamma\right]^{-1}~,
\nonumber\\
\tau_{n,p}=\hbar/\Gamma_{n,p}~,
\end{eqnarray}
respectively. Here $\sigma_{nA}$ and $\sigma_{\gamma A}$ are the
corresponding cross sections, $v_{nA}$ and $v_{\gamma A}$ are the relative
(invariant) velocities, and $\Gamma_{n,p}$ is the neutron (proton)
decay width.


In our calculations for $\sigma_{nA}$ we use the geometrical neutron--nuclear 
cross sections, that is a good approximation for the considered range of 
temperatures ($T\approx (0.5\div 10)$ MeV). The photo--nucleus cross section 
$\sigma_{\gamma A}$ was taken phenomenologically under assumption that it is 
dominated by the giant dipole resonance. These parametrizations of neutron 
and photon cross-sections are in a good agreement with experimental data 
(see discussion in ref.~\cite{SMM}). 
The evaporation decay widths were calculated according to the Weisskopf 
evaporation model as described in ref.~\cite{Botvina87}. Our estimates 
show that at temperatures and densities of interest these reaction times 
vary within the range from 10 to 10$^6$ fm/c, that is indeed very short 
time scale compared to the characteristic
hydrodynamic time of a supernova explosion, about 100 ms \cite{Janka}. 
The nuclear statistical equilibrium is a reasonable approximation 
under these conditions. 

We have calculated the reaction rates of Eq.~(\ref{rate}) for nuclei with 
$A=$60 and $Z=$24, which are typical for stellar nucleosynthesis in dense 
matter and at a typical electron fraction (i.e., the ratio of electron and 
baryon densities $\rho_e/\rho_B$) $Y_e \sim 0.4$. They are presented in 
Fig.~2 as function of neutron density for several temperatures. 
By analyzing this figure one should take into account that the neutron 
density $\rho_n$ is usually by 2--5 times smaller than $\rho$. 
One can see clearly that for densities $\rho_n > 10^{-5}\rho_0$ and for the 
expected temperatures of the environment, $T \loo 5$ MeV, we obtain
$\tau_{\gamma A} >> \tau_{\rm cap}, \tau_{n,p}$, i.e. the
photodisintegration is more slow than other processes.
There exists a range of densities and
temperatures, for example, $\rho_n \goo 10^{-5}\rho_0$ at $T=1$ MeV,
$\rho_n \goo 10^{-3}\rho_0$ at $T=3$ MeV, and $\rho_n \goo 10^{-2}\rho_0$ 
at $T=5$ MeV, where the neutron capture dominates, i.e. 
$\tau_{\rm cap} < \tau_{n,p}$. Under these conditions new channels for 
production and decay of nuclei will appear (e.g. a fast break-up with 
emission of $\alpha$-particles or heavier clusters) which restore 
the detailed balance. We expect that in this situation an  ensemble of 
various nuclear species will be in chemical equilibrium like in a 
liquid-gas coexistence region, as also observed in the multifragmentation 
reactions. Here the nuclear system is fully characterized by the 
temperature $T$, density $\rho$ (which is nearly the same as baryon 
density $\rho_B$), and electron fraction $Y_e$. 
One may expect that modifications of nuclear properties come into force in 
this environment, because of intensive interaction between clusters. 
This is complementary to the well known effects in isolated nuclei: at high 
temperature the masses and level structure in 
hot nuclei can be different from those observed in cold nuclei
(see, e.g., ref.~\cite{Ignat}).

The weak interaction reactions are much slower. The direct and inverse
reactions in Eq.~(\ref{enu}) involve both free nucleons and all nuclei
present in the matter.
It is most likely that at early stages of a supernova explosion
neutrinos/antineutrinos are trapped inside
the neutrinosphere around a protoneutron star \cite{Prakash}.
In this case we should impose the lepton number conservation condition by
fixing the lepton fraction $Y_L$.
Then one should take into account the continuous neutrino flux out of the 
surface of the neutrinosphere propagating through the hot
bubble. Due to large uncertainties in the weak interaction rates,
below we consider three physically distinctive situations:

1) fixed lepton fraction $Y_L$ corresponding to a $\beta$-equilibrium
with trapped neutrinos inside the neutrinosphere (early stage of the 
explosion);

2) fixed electron fraction $Y_e$ but no $\beta$-equilibrium inside a
hot bubble (early and intermediate times);

3) full $\beta$-equilibrium without neutrino (the late times of the 
explosion, after neutrino escape).

The second case corresponds to a non-equilibrium situation which may take 
place in the bubble, before the electron capture becomes 
efficient. Actually, this case is considered as basic for calculations 
of nuclear composition in the hot bubble behind the shock. 
Generally, one should keep in mind that weak reactions are 
often out of equilibrium. Our estimates show that their characteristic 
times range from 10 ms to 10 s depending on thermodynamical conditions and 
intensity of the neutrino wind. Therefore, one should specify what kind 
of statistical equilibrium is expected with respect to weak interaction.

\vspace{0.5cm}

{\bf 3. Formulation of the statistical model.} \\


Below we describe supernova matter as a mixture of nuclear species, 
electrons, photons, and perhaps neutrinos in thermal equilibrium. 
For the macroscopic scales 
one can safely apply the grand-canonical approximation. We call this 
model the Statistical Model for Supernova Matter (SMSM). It was first 
proposed in ref.\cite{Botvina04}.

\vspace{0.5cm}

{\bf 3.1 Equilibrium conditions.} \\

Within the SMSM each particle $i$ with baryon number $B_i$, charge $Q_i$ 
and lepton number $L_i$ is characterized by a chemical potential $\mu_i$, 
which can be represented as 
\begin{equation}
\mu_i=B_i\mu_B+Q_i\mu_Q+L_i\mu_L .
\end{equation}
Here $\mu_B$, $\mu_Q$ and $\mu_L$ are three independent chemical
potentials which are determined from the conservation of total baryon
number $B=\sum_iB_i$ electric charge $Q=\sum_iQ_i$ and lepton number
$L=\sum_iL_i$ of the system. Explicitly, the chemical potentials for 
nuclear species $(A,Z)$, electrons $(e^{-}, e^{+})$, and neutrinos 
$(\nu, \tilde{\nu})$ can be expressed as  
\begin{equation} \label{chem}
 \begin{array}{ll}
\mu_{AZ}=A\mu_B+Z\mu_Q~,~\\
\mu_{e^-}=-\mu_{e^+}=-\mu_Q+\mu_L~,~\\
\mu_\nu=-\mu_{\tilde{\nu}}=\mu_L~.
 \end{array}
\end{equation}
These relations are valid also for nucleons, $\mu_n=\mu_B$ and
$\mu_p=\mu_B+\mu_Q$. 

The corresponding conservation laws can be written as
\begin{eqnarray}
\rho_B=\frac{B}{V}=\sum_{AZ}A\rho_{AZ}~,~\nonumber\\
\rho_Q=\frac{Q}{V}=\sum_{AZ}Z\rho_{AZ}-\rho_e=0~,~\\
\rho_L=\rho_e +\rho_{\nu} -\rho_{\tilde{\nu}}=Y_L \rho_B ~.\nonumber
\end{eqnarray}
Here $\rho_e=\rho_{e^-}-\rho_{e^+}$ is the net electron density, 
$Y_L$ is the lepton fraction. The second equation requires that any 
macroscopic volume of the star is electrically neutral.

The lepton number conservation is a valid concept
only if $\nu$ and $\tilde{\nu}$ are trapped in the system within
the neutrinosphere \cite{Prakash}. If they escape freely from the system,
the lepton number conservation is irrelevant and $\mu_L=0$.
In this case two remaining 
chemical potentials are determined from the conditions of baryon 
number conservation and electro-neutrality. 
Since the $\beta-$ equilibrium may not be achieved in a fast explosive 
process, we also often fix the electron fraction $Y_e=\rho_e / \rho_B$ in 
this case.

\vspace{0.5cm}

{\bf 3.2 Ensemble of nuclear species.} \\

Our treatment of nuclear reactions is based on the statistical 
multifragmentation model 
(SMM) \cite{SMM}, which was very successfully applied for description of 
experimental data. 
For describing an ensemble of nuclear species under supernova conditions
we use the Grand Canonical version of the SMM \cite{Botvina85}. 
After integrating out translational degrees of freedom one can represent 
pressure of nuclear species as
\begin{eqnarray} \label{naz}
P_{\rm nuc}=T\sum_{AZ}\rho_{AZ}
\equiv T\sum_{AZ}g_{AZ}\frac{V_f}{V}\frac{A^{3/2}}{\lambda_T^3}
{\rm exp}\left[-\frac{1}{T}\left(F_{AZ}-\mu_{AZ}\right)\right]~,
\end{eqnarray}
where $\rho_{AZ}$ is the density of nuclear species
with mass $A$ and charge $Z$.
Here $g_{AZ}$ is the ground-state degeneracy factor of species $(A,Z)$,
$\lambda_T=\left(2\pi\hbar^2/m_NT\right)^{1/2}$ is the nucleon
thermal wavelength, $m_N \approx 939$ MeV is the average nucleon mass.
$V$ is the actual volume of the system, and $V_f$ is so called
free volume, which accounts for the finite size of nuclear species.
We assume that all nuclei have normal nuclear density
$\rho_0$, so that the proper volume of a nucleus with
mass $A$ is $A/\rho_0$. At low densities the finite-size effect may be
included via the excluded volume approximation
$V_f/V \approx \left(1-\rho_B/\rho_0\right)$.
We emphasize that this last approximation is commonly accepted in 
statistical models. As we know from nuclear multifragmentation studies 
information about free volume can be extracted from analysis of experimental 
data \cite{EOS}. At densities $\rho_B > 0.1 \rho_0$ the extracted 
$V_f$ may slightly deviate from above approximation, however, they are in 
qualitative agreement. In the present work we remain in the 
framework of the conventional statistical approach, although allowing 
for modifications of nuclear properties. 

The internal excitations of nuclei play an important role in regulating 
their abundance, since they increase significantly their entropy. 
Some authors (see, e.g., ref.~\cite{Japan}) limit the excitation 
spectrum by particle-stable levels known for low excited nuclei. Within the 
SMM we follow quite different philosophy. Namely, we calculate internal 
excitation of nuclei by assuming that they have the same internal 
temperature as the surrounding medium. In this case not only 
particle-stable states but also particle-unstable states will contribute 
to the excitation energy and entropy. This can be justified by the 
dynamical equilibrium of nuclei in hot environment, and supported by 
numerous comparisons with experiment (see part {\bf 2.1}). 
Moreover, in the supernova environment both the excited states and the 
binding energies of nuclei will be strongly affected by the surrounding 
matter. By this reason, we find it more appropriate to use an approach 
which can easily be generalized to include in-medium modifications.
Namely, the internal free energy of species $(A,Z)$ with $A>4$ is
parameterised in the spirit of the liquid drop model, which has been 
proved to be very successful in nuclear physics \cite{Bohr}:
\begin{equation}
F_{AZ}(T,\rho)=F_{AZ}^B+F_{AZ}^S+F_{AZ}^{\rm sym}+F_{AZ}^C~~.
\end{equation}
Here the right hand side contains, respectively, the bulk,
the surface, the symmetry and the Coulomb terms. The first three terms
are taken in the standard form \cite{SMM},
\begin{eqnarray}
F_{AZ}^B(T)=\left(-w_0-\frac{T^2}{\varepsilon_0}\right)A~~, \\
F_{AZ}^S(T)=\beta_0\left(\frac{T_c^2-T^2}{T_c^2+T^2}\right)^{5/4}A^{2/3}~~,\\
\label{fres}
F_{AZ}^{\rm sym}=\gamma \frac{(A-2Z)^2}{A}~~,
\end{eqnarray}
where $w_0=16$ MeV, $\varepsilon_0=16$ MeV, $\beta_0=18$ MeV, $T_c=18$
MeV and $\gamma=25$ MeV are the model parameters which are extracted
from nuclear phenomenology and provide a good description of
multifragmentation data \cite{SMM,ALADIN,EOS,MSU,INDRA,FASA,Dag}. 
However, these parameters, especially the symmetry coefficient $\gamma$, 
can be different in hot nuclei at multifragmentation conditions, 
and they should be determined from corresponding experimental 
observables (see discussion in refs. \cite{Botvina06,bulk,traut}). 

In the electrically-neutral environment the nuclear Coulomb term should be 
modified to include the screening effect of electrons. Within the 
Wigner-Seitz approximation with constant electron density it can be 
expressed as  
\begin{eqnarray}
F_{AZ}^C(\rho)=\frac{3}{5}c(\rho)\frac{(eZ)^2}{r_0A^{1/3}}~~,\\
c(\rho)=\left[1-\frac{3}{2}\left(\frac{\rho_e}{\rho_{0p}}\right)^{1/3}
+\frac{1}{2}\left(\frac{\rho_e}{\rho_{0p}}\right)\right]~,\nonumber
\end{eqnarray}
where $r_0=1.17$ fm and $\rho_{0p}=(Z/A)\rho_0$ is the proton
density inside the nuclei. The screening function $c(\rho)$ is 1
at $\rho_e=0$ and 0 at $\rho_e=\rho_{0p}$. 
Here one can also use an approximation $\rho_e/\rho_{0p}=\rho_B/\rho_0$, 
as in ref.~\cite{Lamb}. In our calculations, we have checked 
that these two choices lead to very similar results, especially at small 
densities. We want to stress that both the reduction of the 
surface energy due to the finite temperature and the reduction of the
Coulomb energy due to the finite electron density favour the formation
of heavy nuclei. Nucleons and light nuclei $(A \leq 4)$ are considered
as structure-less particles characterized only by exact masses and proper 
volumes \cite{SMM}. Their Coulomb interaction is taken into account 
within the same Wigner-Seitz approximation. 

As follows from Eq.~(\ref{naz}), the fate of heavy nuclei depends
strongly on the relationship between $F_{AZ}$ and $\mu_{AZ}$.
In order to avoid an exponentially divergent contribution to the baryon
density, at least in the thermodynamic limit ($A \rightarrow \infty$),
inequality $F_{AZ}\goo \mu_{AZ}$ must hold. 
The equality sign here corresponds to the situation when a big (infinite) 
nuclear fragment coexists with the gas of smaller clusters \cite{Bugaev}. 
When $F_{AZ}>\mu_{AZ}$, only small clusters with nearly exponential mass
spectrum are present. However, there exist a region of thermodynamic 
quantities corresponding to $F_{AZ}\approx\mu_{AZ}$ when the mass 
distribution of nuclear species is a power-law $A^{-\tau}$ with 
$\tau \approx 2$. The advantage of our approach 
is that we consider all the fragments present in this transition region, 
contrary to the previous calculations \cite{Lamb,Lattimer}, which consider 
only one ``average'' nucleus characterizing the liquid phase.

\vspace{0.5cm}

{\bf 3.3 Electromagnetic and weak processes.}\\

At $T,\mu > m_e$ the pressure of the relativistic electron-positron gas 
can be written as
\begin{eqnarray}
P_e=\frac{g_e\mu_e^4}{24\pi^2}\left[1+2\left(\frac{\pi T}{\mu_e}\right)^2+
\frac{7}{15}\left(\frac{\pi T}{\mu_e}\right)^4-\frac{m_e^2}{\mu_e^2}
\left(3+\left(\frac{\pi T}{\mu_e}\right)^2\right)\right] ,
\end{eqnarray}
where first-order correction ($\sim m_e^2$) due to the finite electron 
mass is included, $g_e$=2 is the spin degeneracy factor. 
The net number density $\rho_e$ and entropy density $s_e$ can be
obtained from standard thermodynamic relations, 
$\rho_e=\partial P_e/\partial \mu_e$, and $s_e=\partial P_e/\partial T$,  
which give 
\begin{eqnarray} \label{eden}
\rho_e=\frac{g_e}{6\pi^2}\left[\mu_e^3+\mu_e\left(\pi^2T^2-
\frac{3}{2}m_e^2\right)\right]~,\\
s_e=
\frac{g_e T\mu_e^2}{6}\left[1+\frac{7}{15}\left(\frac{\pi T}{\mu_e}\right)^2-
\frac{m_e^2}{2\mu_e^2}\right]~. 
\end{eqnarray}
Electron neutrinos and antineutrinos are taken into account in the same way, 
but as massless fermions, and with the degeneracy factor twice smaller than 
for the electrons, i.e., $g_{\nu}$=1. 
The photons are always close to the thermal equilibrium, 
and they are treated as massless Bose gas with zero chemical potential. The 
corresponding density 
$\rho_{\gamma}$, energy density $e_{\gamma}$, pressure $P_{\gamma}$, and 
entropy density $s_{\gamma}$ of photons gas are given by standard formulae: 
\begin{equation}
\rho_{\gamma}=\frac{g_{\gamma}\xi(3) T^3}{\pi^2 \hbar^3 c^3}~,~
e_{\gamma}=\frac{g_{\gamma}\pi^2 T^4}{30 \hbar^3 c^3}~,~
P_{\gamma}=\frac{e_{\gamma}}{3}~,~
s_{\gamma}=\frac{4 e_{\gamma}}{3 T},
\end{equation}
where $g_{\gamma}$=2. 

All kinds of particles (nuclei, baryons, electrons, neutrinos, photons) 
contribute 
to the free energy, pressure and other thermodynamical characteristics of 
the system, and we sum up all these contributions. Within the model we 
calculate densities of all particles self-consistently 
by taking into account the relations between their 
chemical potentials.

\vspace{0.5cm}

{\bf 3.4 Comparison with the Lattimer-Swesty model.}\\

We have performed calculations for sets of physical conditions expected 
during the collapse of massive stars and subsequent supernova explosions. 
We take baryon number $B=$1000 and perform calculations 
for all fragments with 1$\leq A \leq$1000 and 0$\leq Z \leq A$
in a box of fixed volume $V$. This volume is determined by the average 
baryon density $\rho_B=B/V$. This restriction on the size of 
nuclear fragments is fully justified in our case, since 
fragments with larger masses ($A>$1000) can be produced only at
very high densities $\rho \goo 0.5\rho_0$ \cite{Bethe,Lamb}, which are 
appropriate for the regions deep inside the protoneutron star, and
which are not considered here.

In the beginning we compare our model with calculations within 
other models on the marked. It is necessary to mention that all 
models treat electrons and photons in the same way, therefore, differences 
appear entirely due to different description of nuclear species. 
One can expect that in the case of domination of radiation processes the 
results will be similar. 


Most supernova simulations are performed with the equation of state of 
Lattimer--Swesty (LS) \cite{Lamb,Lattimer}, where an ensemble of heavy 
nuclei is replaced by a single ``average'' nucleus. Therefore, one can 
compare only integral characteristics of the stellar matter. 
In Fig.~3 we compare our SMSM results with the LS 
model. We show entropies, pressure, mass fractions of alpha particles 
($X_{alpha}$), and heavy fragments ($X_{heavy}$), at different temperatures, 
and at a fixed electron fraction, versus densities. 
In the SMSM $X_{heavy}$ includes all fragments with $A>$4, whereas in 
the LS model it is only  the share of the ``average'' heavy nucleus. 
The LS calculations 
were taken from ref. \cite{Janka2}. One can see that the average 
thermodynamical characteristics (pressure, entropy) are very close 
in the two models. This remains also true if we extend comparison to 
other models, e.g., reported in \cite{Janka2,Japan}. 
However, as seen from Fig.~3, mass fractions of nuclei are very different. 
As was mentioned 
in \cite{Janka2}, a small yield of alpha clusters in the LS model may 
be caused by mistakes in calculations of the Coulomb corrections to 
their binding energies. We stress again that the details of the nuclear 
composition 
are very important for dynamics of the explosion, since it influences 
the total energy balance, and determines the weak reaction rates. 

\vspace{0.5cm}

{\bf 4. Composition of matter.}\\

In this section we present the SMSM results concerning 
nuclear and lepton composition of stellar matter, which are important 
for determining general thermodynamical characteristics of the matter, 
such as energy density, pressure, and entropy. The 
energy deposition into the matter with photons and 
neutrinos produced by external sources is also considerably influenced by 
this composition. 
We pay special attention to the contribution of nuclear species whose 
properties may be modified in dense environment as follows from recent 
findings in nuclear multifragmentation reactions. 

\vspace{0.5cm}

{\bf 4.1 Electron and neutron fractions of stellar matter.}\\

Within the SMSM we can calculate the electron fraction in the 
electrically-neutral matter under assumption of full $\beta$-equilibrium. 
The appropriate astrophysical sites, where this may happen, are 
the relatively slow collapse stage, and the very late stages of 
the explosion, after cooling down the matter and neutrino escape. 
In this case we calculate self-consistently densities of all species 
by using relations between their chemical potentials, Eq.~(\ref{chem}), 
with $\mu_L$=0. The net electron density, which is equal to the proton 
density, is explicitly given as a function of the chemical potential 
in Eq.~(\ref{eden}). 
Fig.~4 presents the fractions of free electrons and neutrons as functions 
of baryon density. 


On the left panels the results are shown for the $\beta$-equilibrium 
neutrino-less matter at $T=$1 and 3 MeV. 
One can clearly see two general trends: with increasing baryon density 
the electron fraction $Y_e$ gradually decreases and the neutron 
fraction $Y_n$ increases. 
At small densities, which 
correspond to low $\mu_e$, the electrically-neutral matter tends to be 
isospin symmetric, with a large amount of electrons. 
In the case of low temperatures ($T\loo 1$ MeV) the protons are captured 
in most bound nuclei with $A\sim$ 50--60. 
As was realized long time ago \cite{Bethe}, at large 
densities the electrons are absorbed by protons in the inverse 
$\beta$-decay process, that is driven by high electron chemical potential. 
When we increase temperature ($T \goo 3$ MeV) the nuclei dissociate into 
protons and neutrons, that helps to capture electrons at large densities 
also. At the same time, the number of free neutrons 
increases rapidly at higher baryon densities for both low and high 
temperatures. This is important for maintaining a high rate of nuclear 
reactions to generate equilibrium ensemble of nuclei. 
At low temperatures ($T \sim 1$ MeV) a noticeable change in the trend is 
seen below $\rho_B\approx 10^{-4}\rho_0$. At these densities 
most neutrons are bound in large nuclei, which are still present in the 
matter. For example, $Y_n \approx$ 0.2 means that 80\% of neutrons are
trapped in the nuclei. At even lower densities, when heavy nuclei 
disappear (see Fig.~3) the number of free neutrons increases. The same 
trend at higher densities is explained by the fact that more and more 
neutrons are dripping out of nuclei, since the matter contains less and 
less protons. For example, at $\rho_B>10^{-3}\rho_0$ more 
than half of the neutrons are free. This behavior correlates with decreasing 
the number of electrons and a relatively small share of heavy nuclei in the 
system. At higher densities, the structure of matter may change because of 
the neutrino/antineutrino and electron/positron capture reactions 
\cite{Bethe}. 

The right panel of Fig.~4 shows the results obtained under the condition 
of lepton number conservation, i.e., at fixed values of lepton fractions 
$Y_L$=0.1, 0.2, 0.3. These values are 
consistent with uncertainties concerning the neutrinosphere radius 
discussed in literature. One can see that in this case the number of free 
neutrons always drops with density reflecting formation of very big nuclei 
and transition to the liquid phase at $\rho_B \rightarrow \rho_0$. While 
the electron fraction stays nearly constant, exhausting around 80--90\% 
of the total $Y_L$. These results show that weak reactions affect 
significantly the composition of supernova matter.

\vspace{0.5cm}

{\bf 4.2 Mass fractions of light and heavy nuclei.}\\

As well known, at low densities and temperatures the nuclear matter exists 
in the form of isolated nuclei and nucleons. At terrestrial conditions 
the nuclei capture electrons and become atoms. However, at supernova 
conditions the atoms are fully ionized, therefore, the nuclei are embedded 
in more or less uniform background of electrons and neutrons. This 
surrounding to a large extent determines the nuclear composition of 
stellar matter. 

Figures 5 and 6 demonstrate the mass fractions of nuclear matter contained 
in heavy nuclei (with mass numbers $A > 4$), $\alpha$-particles, neutrons 
and protons, for different electron fractions $Y_e$. One can see that at 
low temperatures ($T < 1$ MeV) the matter is mainly composed of heavy nuclei. 
If the share of electrons is small, the free neutrons are also present. With 
increasing temperatures the heavy nuclei gradually disintegrate into 
$\alpha$'s, neutrons and protons. At low densities this disintegration 
happens already at moderate temperatures $T \sim$ 1--2 MeV, while 
at subnuclear densities ($\rho \sim 0.1 \rho_0$) the heavy nuclei 
survive even at higher temperatures, though they become very excited. One 
should bear in mind, however, that in this case we are dealing with the 
dynamical equilibrium between decay of excited nuclei and absorption of 
surrounding nucleons, as regulated by the reaction rates presented 
in Fig.~2. An interesting observation is that heavy nuclei first break-up 
into light clusters (like $\alpha$) and then these clusters dissolve into 
nucleons with temperature. This is clear seen in the 
yields of $\alpha$--particles, which demonstrate a 'rise and fall' 
behaviour both with increasing temperature and decreasing density.
As seen from Figs.~7 and 8, small clusters with $A$=2 and 3 are also 
produced in the transition region. 



\vspace{0.5cm}

{\bf 4.3 Nuclear mass distributions.}\\

The properties of heavy nuclei are very important for understanding 
processes taking place in stellar matter. 
In the bottom panel of Fig.~7 we show mass distributions of nuclear 
species at $\rho_B = 10^{-3} \rho_0$ and several temperatures. 
At low temperatures the distribution of heavy nuclei looks like a Gaussian 
with a well defined peak (see also ref.~\cite{Botvina04}). In this case the 
average thermodynamic characteristics of the system may not be much 
different from the ones calculated under assumption of an ``average'' 
nucleus as in ref.~\cite{Lattimer}, see Fig.~3. 
However, even in this case, the width of the distributions may be important 
for calculations of weak reactions in matter. By increasing temperature 
we move into a coexistence region of the nuclear liquid-gas phase 
transition: The mass distributions become 'U-shaped', and they 
contain all nuclei from light to heavy ones. At higher temperatures the mass 
distributions have exponential shape. These distributions can not 
be even approximately characterized by an ``average'' nucleus. 
This evolution of fragment mass distributions is well established 
in nuclear multifragmentation reactions \cite{SMM,Dag}. 

The average charge-to-mass ratio for all nuclei is demonstrated in the top 
panel of Fig.~7. For this quantity both the symmetry and Coulomb energies of 
fragments are crucially important. The charges of heavy fragments show 
rather regular behaviour: The $Z/A$ ratio decreases slowly with mass number 
because of Coulomb interaction. It decreases less rapidly than in 
multifragmentation reactions due to the screening effect of electrons. 
From our calculations we came to the conclusion that the charge 
distribution of nuclei at given $A$ can be approximated by a Gaussian, 
with the width 
$\sigma_Z \approx \sqrt{AT/8\gamma}$ \cite{Botvina87,Botvina85,Botvina01}, 
where $\gamma$ is the coefficient in the symmetry energy 
(see Eq.~(\ref{fres})). 


We have found that the phase transition from heavy nuclei to light 
fragments (nuclear gas) always proceeds through the same sequence of 
mass distributions: 
'U-shape', power-law, and 
exponential ones, both with increasing temperature and 
decreasing density. This opens the possibility to study the critical 
behaviour in stellar matter, in the same way as was previously done 
in multifragmentation reactions \cite{EOS,Dag}. 
One can find examples of the temperature-driven transitions in 
refs.~\cite{Botvina04,Botvina05} and in Fig.~7 (see also Fig.~22). 
In Fig.~8 we demonstrate an example of the density-driven transition 
at fixed temperature of 1 MeV. As we can clearly see, the mass distributions 
evolve from 'U-shape' at densities (in units $10^{-5}\rho_0$) 1.0 and 
0.32 to power-law at 0.18 and 0.1, and finally to exponential at 0.03. 
At density 0.18 the distribution of large clusters is most flat, 
that may be considered as 
a critical point of the phase transition \cite{Bugaev}.

\vspace{0.5cm}

{\bf 4.4 Shell effects.}\\


Up to now we have used the liquid-drop description of nuclei in the 
stellar matter. However, it is well known that at low temperatures 
($T\loo 1$ MeV) the shell corrections to nuclear masses becomes important, 
at least, at terrestrial densities of matter. 
It is likely that the shell effects may also play a role in stellar matter 
at subnuclear densities. In this section we demonstrate how shell effects 
may influence the mass and charge distributions of nuclei. 
In particular, we have analyzed possible existence of superheavy elements 
in stellar matter, assuming that they are sufficiently long-lived. 
It was assumed that there is a certain shell correction to the free energy 
$F_{AZ}$ of a specific nucleus, $\Delta F_{AZ}$, taken as follows:
\begin{equation}
\Delta F_{AZ}=-\left[E_{sh}exp \left(-\frac{(N-N_{sh})^2}{2\sigma_N^2}\right)+
            E_{sh}exp \left(-\frac{(Z-Z_{sh})^2}{2\sigma_Z^2}\right)\right].
\end{equation}
We have assumed existence of the island of stability around $Z_{sh}$=120 
and $N_{sh}$=180 \cite{Greiner}. The maximum energy of the shell correction 
is chosen to be $E_{sh}$=5 MeV, and 
the widths of the shell are $\sigma_{N}= \sigma_{Z}=$5 MeV. These values 
are consistent with the magnitudes of the shell effects discussed for 
superheavy elements. In Fig.~9 we compare the calculated charge and 
mass distributions without and with the shell corrections (i.e., without and 
with the term $\Delta F_{AZ}$). We consider typical conditions in supernova 
matter where the production of superheavy nuclei is still possible, i.e., 
$T=$1 MeV, $\rho_B \approx 0.05\rho_0$, $Y_e=0.2$. One can see that 
a pure liquid-drop description predict Gaussian-like mass and charge 
distributions of fragments which move to smaller values with decreasing 
density. When the shell constraint is imposed, the yields become essentially, 
by factor 2--5, larger in the vicinity of neutron ($N_{sh}$) and proton 
($Z_{sh}$) 'magic' numbers. Moreover, because of the increased binding, 
these magic nuclei are abundantly produced in a rather broad region of 
density. For example, the neutron shell of $N_{sh}$=180 dominates clearly 
at densities of both $0.03\rho_0$ and $0.05\rho_0$. 


This result makes possible to discuss nucleosynthesis of 
superheavy elements at supernova conditions. As has been already 
pointed out in ref.~\cite{Botvina04}, there is a chance that heavy nuclei 
could be produced at subnuclear densities, and then ejected into 
the space. At considered small temperature (1 MeV) the effect of 
secondary deexcitation will be minimal, only few nucleons will be lost. 
The fission channels for these neutron-rich nuclei may also be suppressed 
by the shell effects \cite{Nix} and electron screening \cite{Burven}. 
However, one can expect that these nuclei will fastly emit neutrons above the 
neutron drip-line, undergo abundant $\beta$ decay and, possibly, $\alpha$ 
emission. We are planning to analyze all these processes in the forthcoming 
publications \cite{s-heavy}. 
Here we mention only a possible scenario how such superheavy nuclei can be 
ejected into space. Since the synthesis of heavy and 
superheavy nuclei is only possible at rather high baryon density 
$\rho_B \sim 0.05\rho_0$ it is most likely that such nuclei will not be 
ejected in the course of the supernova explosion. Instead, they will be 
accumulated at the surface of a newly produced neutron star. If this star 
is in a binary system with another neutron star, white dwarf, or even a 
black hole, there is a chance of their collision at later stages of 
evolution.  Then a part 
of the stellar material will be ejected in space, while the other part 
may collapse into a black hole. 
One can also speculate about asymmetric explosions of supernovas, 
acceleration of nuclei by the neutrino wind, and starquakes which may provide 
this ejection 
(see, e.g., refs. \cite{star-collide,neutrino-wind,star-quakes}). 
The search for new mechanisms of nucleosynthesis is motivated by the fact 
that the traditional $s$- and $r$-processes have serious problems 
to explain synthesis of fissioning nuclei larger than lead \cite{Qian}. 

\vspace{0.5cm}

{\bf 5. Thermodynamical characteristics of stellar matter.}\\

In this section we present general thermodynamical characteristics of 
stellar matter, such as energy density, pressure and entropy as functions 
of temperature $T$, baryon 
density $\rho_B$, electron fraction $Y_e$, as well as the nuclear 
composition. In hydrodynamical 
simulations most important role is played by Equation of State (EOS), 
which connects pressure with the energy density. 
The dynamics of collapse and explosion depends essentially 
on the EOS \cite{Janka}. 

\vspace{0.5cm}

{\bf 5.1 Caloric curve, pressure and entropy}\\


One of the main inputs for dynamical simulations of supernova explosions 
is the total thermal energy deposited in the matter. 
In Fig.~10 we show so-called caloric curves, i.e., the thermal energy 
per nucleon of the matter as a function of temperature. For convenience, 
the actual energy per nucleon is shifted by the value of 16 MeV, which 
corresponds to the bulk binding energy of nuclear matter at 
$\rho_B = \rho_0$. Our calculations show that nuclear contributions 
dominate at high densities and low temperatures, where heavy nuclei 
survive. Due to the internal excitation of these nuclei according to the 
compound nucleus law, $E^{*} \sim T^2$, the caloric curve has a parabolic 
shape at $\rho_B \sim 0.1-0.001 \rho_0$. However, at low densities and 
high temperatures electrons and photons dominate, and the caloric curve 
behaves according to the Stefan--Boltzmann law $E^{*} \sim T^4$, i.e., 
the energy per nucleon grows very rapidly with temperature. 
One can also see that the nuclear contributions become nearly independent on 
baryon density at high temperatures. Namely, they approach the Boltzmann 
limit, $E^{*} \sim 1.5T$, when nuclear matter disintegrate completely into 
nucleons. It is instructive to note that at high densities 
$\rho \sim 0.1\rho_0$ and proton fractions $Y_p=Y_e \sim 0.4$ which are 
typical for normal 
nuclei, the caloric curve is determined mainly by the nuclear species, and 
it reminds very much the caloric curves extracted from the 
multifragmentation reactions \cite{SMM}. In this case, the excitation 
energy increases rapidly at temperatures $T \approx 4-6$ MeV, which 
correspond to the maximum in the heat capacity. This is a characteristic 
feature of the liquid-gas phase transition \cite{Dag,Bugaev}. 


Pressure is another important characteristic of the matter, 
which, in competition with the gravitational pressure, determines the 
structure and dynamics of the stellar system. 
In Figs.~11 and 12 we plot the pressure versus baryon density for different 
temperatures. There are several important features to be mentioned. First, 
the nuclear contributions are mainly important in the intermediate 
density region $(10^{-4}-10^{-2})\rho_0$, where, depending on 
temperature, more and more free nucleons are present in the nuclear matter. 
In the case of full disintegration of nuclear species into nucleons, 
the nucleons may contribute up to 50\% to the total pressure. Second, at 
higher densities the pressure is dominated by the relativistic electrons, 
since their chemical potential becomes very high. Third, at very low 
densities and high temperatures the radiation pressure dominate, which is 
proportional to $T^4$ and does not depend on baryon density. One can see 
that the nuclear pressure is higher in the case of a low electron 
fraction ($Y_e$=0.2), because of a considerable abundance of free neutrons, 
even if large nuclei are present in the matter. 
In the modern hydrodynamical simulations of supernova explosions the shock 
stalls at densities around $10^{-6}-10^{-5}\rho_0$ \cite{Janka}. 
This happens partly because a significant fraction of shock energy is 
used for disintegration of infalling nuclei, from C to Fe. 
Therefore, survival of medium and heavy nuclei would contribute 
essentially to the revival of the shock. 


The entropy per baryon $S/B$, which is shown in Fig.~13 as a function of 
temperature, is an important characteristic of the exploding matter. 
One can notice that it correlates strongly with behavior of the caloric curve. 
At low temperatures and high densities the nuclear contribution to the 
entropy dominates. At high temperatures, the nuclei disintegrate into 
nucleons and the nuclear entropy depends only logarithmically 
on temperature and density according to the Boltzmann gas law. 
Usually, at entropy greater than 10 units per nucleon only nucleon gas 
without heavy nuclei is present in the system. The main contribution to the 
total entropy in this case is provided by the radiation and 
electron-positron pairs. This contribution does not depend on density and 
is proportional to $T^3$. The total entropy has a jump across the shock, 
which can be explained in part by disintegration of heavy nuclei. 
Therefore, even small differences in nuclear properties in medium, 
in comparison with isolated nuclei, may lead to significant effects 
in the shock dynamics. 


\vspace{0.5cm}

{\bf 5.2 Adiabatic trajectories.}\\

Some important processes in stellar matter proceed with approximately 
constant entropy. For example, the collapse of a massive star before 
re-bounce is characterized by the entropy around one unit per nucleon. 
After propagation of the shock wave the entropy increases drastically. 
However, subsequent evolution of the matter is again close to isentropic, 
and this assumption is quite valid for nucleosynthesis. For this 
reason we consider adiabatic trajectories in the $T - \rho_B$ plane, 
and calculate fragment mass distributions and thermodynamical functions 
along these trajectories. 


In Fig.~14 we show adiabatic trajectories for several fixed S/B. This 
representation of the phase diagram on $T - \rho$ plane is very convenient 
for understanding thermodynamical properties of stellar matter (compare also 
with Fig.~1). 
It is also possible to make a rough estimation of nuclear composition of 
the matter: As well known from previous studies \cite{Bethe}, at S/B=1 
many heavy nuclei exist in the matter, while at the entropy as high as 
$S/B$=20 the baryonic matter consist mainly of free nucleons and hadron 
resonances. Our calculations confirm this expectation. 


It is instructive to compare mass fractions of different 
nuclear species along the adiabats for different thermodynamical conditions, 
in order to get an idea about evolution of nuclei during the whole 
collapse--explosion process.  In Fig.~15 we show these fractions (similar to 
Fig.~6) for S/B=1, where the mass fraction of heavy nuclei is close 
to 1, and for S/B=8, where the nuclei undergoes deep disintegration. In the 
latter case the fraction of nuclei with $A>4$ is essential only at high 
densities, and there are no large fragments at low densities. At S/B=8 the 
fraction of $\alpha$ particles has a very interesting behavior: It has a 
minimum at intermediate densities ($\rho_B \sim 10^{-3}\rho_0$), but then 
it increases with decreasing density, and completely dominates at low 
densities. For this fixed entropy 
a large binding energy of $\alpha$ particles is very important 
for the thermodynamical balance in the system. Therefore, $\alpha$ 
particles can be preferable 
instead of individual nucleons at the low density conditions. 
The production of $\alpha$ clusters may help to revive the shock wave 
by maintaining a sufficiently high temperature behind the shock.


The mass distributions of nuclei evolve strongly along the isentropic 
trajectories. In Fig.~16 we demonstrate this evolution in stellar matter 
with a large electron fraction $Y_e =0.4$ for two cases 
corresponding to a low entropy S/B=1, and to a higher entropy S/B=4, 
where the contribution of heavy fragments is still essential. It is 
important that in the first case (S/B=1) at high baryon density 
($\rho_B = 0.1\rho_0$) and temperature ($T=$3.39 MeV), the distribution 
of heavy nuclei is centred around $A \sim$130 and has a large width 
$\sigma_A \sim 50$. Therefore, heavy nuclei with mass number higher 
than 200, as well as very small clusters, coexist in the system. 
With decreasing density the distribution of nuclei shifts to smaller 
masses and becomes more narrow. Finally, the distribution moves into  
the iron region, where the binding energy is maximal. At higher entropy 
(S/B=4) we have an opposite situation. The masses of produced nuclei become 
larger with decreasing density. And at very low densities the nuclei reach 
finally the iron region. The reason for this interesting behavior is that 
the temperature drops essentially (from 10 to 1 MeV, as seen from Fig.~14) 
along the adiabatic trajectory. 


All above discussed trends take also place in the case of isospin asymmetric 
matter with a small electron fraction, $Y_e=0.2$, see Fig.~17. Besides of the 
expected effect of increasing number of free neutrons, we obtain here that 
more heavy nuclei are produced at the same densities. In the same time these 
nuclei are very neutron-rich. For example, at S/B=1 and $\rho=0.1\rho_0$, 
for $Y_e=0.2$ we have an average charge of big nuclei 
$\langle Z \rangle /A \approx 0.28$, in comparison with 
$\langle Z \rangle /A \approx 0.41$ for $Y_e=0.4$. 
Because of formation of the unusual heavy nuclei, the temperature is lower 
for small electron fractions. One can conclude from analysis of Figs.~15, 16, 
and 17 that clusterization of nuclear matter at low densities should have 
important consequences for the explosion, at least via the energy balance. 
However, this effect will be even stronger if we include into consideration 
the modifications of the weak reaction rates caused by the clustering (see 
below). 


In Fig.~18 we demonstrate the total adiabatic pressure as function of baryon 
density. As expected, it increases with density, and with entropy. 
One can see a nearly linear relation between ln(P) and ln($\rho$). For a 
fixed entropy this relation is usually expressed as the politropic equation 
\begin{equation}
P \sim \rho^{\Gamma_{ad}},
\end{equation}
where $\Gamma_{ad}$ is an effective adiabatic index. 
In Fig.~19 we show the behavior of $\Gamma_{ad}$ as a function of density 
for two values 
of entropy per baryon. At S/B=1, when heavy nuclei still exist, the 
adiabatic index is nearly constant and close to 4/3, which is expected 
for relativistic electron gas. At S/B=8 the $\Gamma_{ad}$ coefficient shows 
more interesting behavior: it changes considerably and goes through the 
maximum. The maximal value of 1.5 is reached around $10^{-4}-10^{-2}\rho_0$. 
This corresponds to a change of the pressure slope seen in Fig.~18. This 
effect is caused by a nearly complete disintegration of nuclei into nucleons, 
which takes place in this density region (see Fig.~15). The production 
of heavy nuclei at higher densities, and production of $\alpha$ particles 
at lower densities, lead to decreasing $\Gamma_{ad}$ at the both side of 
the maximum. Matter with $\Gamma_{ad} < 4/3$ can not resist the gravity 
and, therefore, unstable with respect to gravitational collapse. On the 
other hand, matter with $\Gamma_{ad} > 4/3$ can provide conditions for 
outward propagation of the shock wave during the supernova explosion. 


Finally, let us consider now the adiabatic sound velocity, 
$c_s^2=\partial P/ \partial \rho |_s$, which plays an important role in 
hydrodynamical simulations. 
For example, when the collective velocity of matter exceeds $c_s$ a shock 
wave is generated. In Fig.~20 we present the sound velocity along 
different adiabates. In this case it can be obtained as 
$c_s^2=\Gamma_{ad}\cdot P/\rho$. As expected, $c_s$ increases 
with density. However, in the case of S/A=8, there is a peak around 
$10^{-2}\rho_0$, which is caused by the same physical reasons 
as the maximum of $\Gamma_{ad}$, shown in Fig.~19. We note that a small 
sound velocity in comparison with the light velocity give a justification 
for using nonrelativistic formulas for baryonic and nuclear degree of 
freedom. 


\vspace{0.5cm}

{\bf 6. Possible in-medium modification of nuclear properties.}\\

{\bf 6.1 Reduction of symmetry energy. } \\

As we have mentioned in section {\bf 3} multifragmentation reactions 
open a unique possibility to investigate clusterization of nuclear 
matter at subnuclear densities. Recently, the symmetry energy of hot 
nuclei was extracted from experimental data 
\cite{Botvina05,LeFevre,Iglio,Souliotis}, 
and it was demonstrated that the $\gamma$ coefficient, see Eq. (\ref{fres}), 
is considerably reduced as compared with the values expected for cold 
isolated nuclei. This effect becomes stronger with increasing excitation 
energy. For example, in Fig.~21 we show the extracted values of the $\gamma$ 
coefficient, which go significantly down with decreasing impact parameter 
$b$, i.e., for more and more central collisions. 
The empirical value of the $\gamma$ coefficient, approximately $25$~MeV, 
was obtained for isolated nuclei from the liquid-drop description of their 
binding energies. As one can see, at high excitation energies it drops 
down to $\approx 15$ MeV, and may even lower, as follows from the 
analysis of ref.~\cite{LeFevre}. As discussed in refs. 
\cite{traut,LeFevre,Iglio,Souliotis,Botvina06,bulk} this change can be 
explained 
by the reduction of the fragment density, modification of the nuclear 
surface energy, and the influence of nuclear environment. 


Now we come back to the comparison of conditions which can be reached 
in multifragmentation reactions and in supernova explosions. Specifically, 
in Fig.~22 we demonstrate the similarity of fragment mass distributions 
for these two physical systems, calculated within the SMM and SMSM using 
the same description of hot fragments in dense environment. 
The density and temperature of stellar matter are chosen close to the ones 
typical for multifragmentation reactions. The electron fraction 
($Y_e=$ 0.2) corresponds approximately to the deleptonization values 
obtained at this density. One can see that the evolution of mass 
distributions with excitation energy and temperature is qualitatively 
similar for both cases (see also section {\bf 4.3} and 
refs.~\cite{SMM,Botvina08}). The transition from 
the 'U-shaped' mass distribution to the exponential one, a characteristic 
feature of the liquid-gas phase transition, is well pronounced in both cases 
too. However, in the supernova environments much heavier and neutron-rich
nuclei can be produced because of screening effect of surrounding electrons. 
In this figure we also demonstrate how important is to extract reliable 
information about the symmetry energy of hot nuclei. 
As one can see from mass yields at 3 MeV per nucleon in top panel, 
changing $\gamma$ coefficient from 25 to 15 MeV has practically no 
effect on the mass distributions of fragments produced in 
nuclear reactions. As was shown in refs. \cite{Botvina06,nihal}, in nuclear 
multifragmentation reactions the most noticeable effect is 
that the isotope distributions become broader at smaller $\gamma$. 
However, the symmetry coefficient $\gamma$ has a dramatic 
influence on masses of nuclei produced in supernova environments. 
One can see from Fig.~22 that much more heavy (and more neutron-rich) nuclei 
can be formed in this case. This effect makes very likely production of 
heavy and superheavy nuclei in supernova environments. 
In the following these hot nuclei should undergo de-excitation, and their 
decay products may either survive on the surface of a neutron star or 
be ejected into inter-stellar space. Also they can 
serve as seeds for subsequent $r-$process, as discussed in section 
{\bf 4.4}. Therefore, 
studying the multifragmentation reactions in the 
laboratory is important for understanding how heavy elements were 
synthesized in the Universe. 


\vspace{0.5cm}

{\bf 6.2 Electron capture rate. } \\

Reduction of the symmetry energy of hot nuclei can be very important
for weak processes. Here we consider a typical example related to
deleptonization of matter, e.g., the electron capture by nuclei. One
should bear in mind that the electron fraction is crucial for dynamics
of stellar matter, since the electron pressure dominates at subnuclear
densities. The calculation of the electron capture rate, $R_e$,
was carried out with the method suggested in ref.~\cite{langanke1}. It is
based on an independent particle model and assumes dominance of Gamow-Teller
transitions. The electron chemical potential $\mu_e$ and the reaction
$Q$-value are the most important energy scales of the capture process. 
It is clear that the 
$Q$-value is directly related to the symmetry energy coefficient $\gamma$.
A good approximation for the capture rate (per second) on an isolated nucleus
is the expression \cite{langanke1}:
\begin{equation}
R_e=\frac{0.693B_g}{t_g}\left(\frac{T}{m_{e}c^2}\right)^5
\left[F_4(\eta)-2\xi F_3(\eta)+\xi^2 F_2(\eta)\right],
\end{equation}
where $t_g=$6146$s$, $B_g=$4.6 represents a typical (Gamow-Teller plus
forbidden) matrix element, $\xi=(Q-\delta E)/T$, $\eta=(\mu_e+Q-\delta E)/T$,
$\delta E=$2.5 MeV, and $F_k$ are the relativistic Fermi integrals of order 
$k$. It is instructive to normalize per nucleon this rate by taking into
account the whole ensemble of heavy nuclei produced in stellar matter:
$\langle R_e \rangle=\sum\rho_{AZ}R_e/\rho_B$.
Figure 23 demonstrates that the electron capture rate in stellar
matter depends very essentially on the symmetry energy of nuclei.
One can see that at relatively high densities $\rho_B \sim 0.1\rho_0$ the
electron capture rate changes only by 20-50\%, if we adopt
the reduced symmetry energy coefficient $\gamma \approx 15$ MeV.
This is because a high electron chemical potential drives the reaction.
However, at small densities (below $10^{-3}\rho_0$), when heavy nuclei 
with large charge still 
exist (at least at low temperatures), the effect of reduced $\gamma$ is
dramatic, of two-three orders of magnitude. We note, that at these relatively
low densities and temperatures the nuclear chemical equilibrium may 
already be problematic \cite{Botvina09}, although some authors keep 
using it in network calculations \cite{network}. We believe that hot nuclei
can interact with each other by neutron exchange in this case. This 
situation is similar to what we have at higher densities of
nuclear matter in multifragmentation reactions. Therefore, the effect of
reduction of the symmetry energy observed in multifragmentation may also
take place in the supernova environments and be responsible for significant 
enhancement of weak reaction rates. 


\vspace{0.5cm}

{\bf   Conclusions}\\

We have formulated a statistical approach (SMSM) designed to describe
supernova matter at subnuclear densities.
It may be applied for a broad variety of stellar processes, including 
the collapse of massive stars and supernova explosions, clusterization 
of nuclear matter in the crust of neutron stars, nuclear composition in 
merging binary stars, etc. 
The model includes the whole ensemble of nuclear species, as well as 
photons and leptons ($e^{-}, e^{+}, \nu, \tilde{\nu}$). The model 
fully accounts for the nuclear liquid-gas phase transition, which was 
previously under active investigation in nuclear reactions. In general, 
we emphasize a close connection of the processes in stellar matter with 
multifragmentation reactions studied in laboratories. 

We have calculated main thermodynamical characteristics of the stellar matter 
under different assumption on the lepton fractions. Nuclear degrees of freedom 
contribute essentially to the energy and the entropy at high densities. 
Whereas, at low densities and high temperatures the photons and leptons 
contributions dominate. The nuclear contribution to pressure becomes essential 
only when nuclei completely dissociate into nucleons. Accordingly, 
the adiabatic index increases considerably in this region. On the other hand, 
we have found that the $\alpha$ particle production at low densities and 
moderate entropies can be an important process, which should be correctly 
taken into account in the dynamical simulations of the explosion. 

The comparison with the Lattimer-Swesty model shows that thermodynamical 
quantities of the matter, e.g., pressure, are not very different in two 
models. The reason is that both models treat the leptons and photons in 
a similar way. Considerable differences appear in the yields of $\alpha$ 
particles and heavy nuclei. We believe that the SMSM provides more 
realistic mass and charge distributions of hot nuclei, without any 
additional constraint on their sizes. 

As a result of our calculations, we especially emphasize the evolutionary 
nature of the mass and charge distributions of produced heavy fragments. 
These distributions carry important information regarding the nuclear 
liquid-gas phase transition in the stellar matter. Also these nuclei 
participate in many processes which determine the energy deposition and 
dynamics of the collapse and explosion. Motivated by recent findings in 
nuclear multifragmentation reactions we have analyzed how possible 
in-medium modifications of nuclear properties, in particular, a reduction 
of the symmetry energy, can influence the fragment 
yields and weak processes. We have found that these effects can be very 
important, e.g., for the electron capture by nuclei, which is responsible for 
deleptonization of matter. At the same time, they can increase the yield of 
big neutron-rich nuclei. We have discussed new mechanisms of 
nucleosynthesis leading to the production of heavy and superheavy nuclei 
in supernova environments. In particular, the shell effects existing at 
relatively low temperatures may provide an additional enhancement factor 
for their formation.

The authors thank W. Trautmann, J. Schaffner-Bielich, M. Hempel, K.
Langanke, Th. Janka, J. Lattimer, M. Liebendoerfer, Th. Buervenich, 
H. St{\"o}cker, 
and W. Greiner for many fruitful discussions. One of us (A.S.B.) 
acknowledges financial support from the Helmholz International Center for 
FAIR. 
This work was partly supported by DFG grant 436RUS 113/711/02 (Germany)
and by grant NS-3004.2008.2 (Russia).

\vspace{0.5cm}

\begin{figure} [tbh]
\vspace{-1cm}
\hspace{2cm}
\includegraphics[width=10cm]{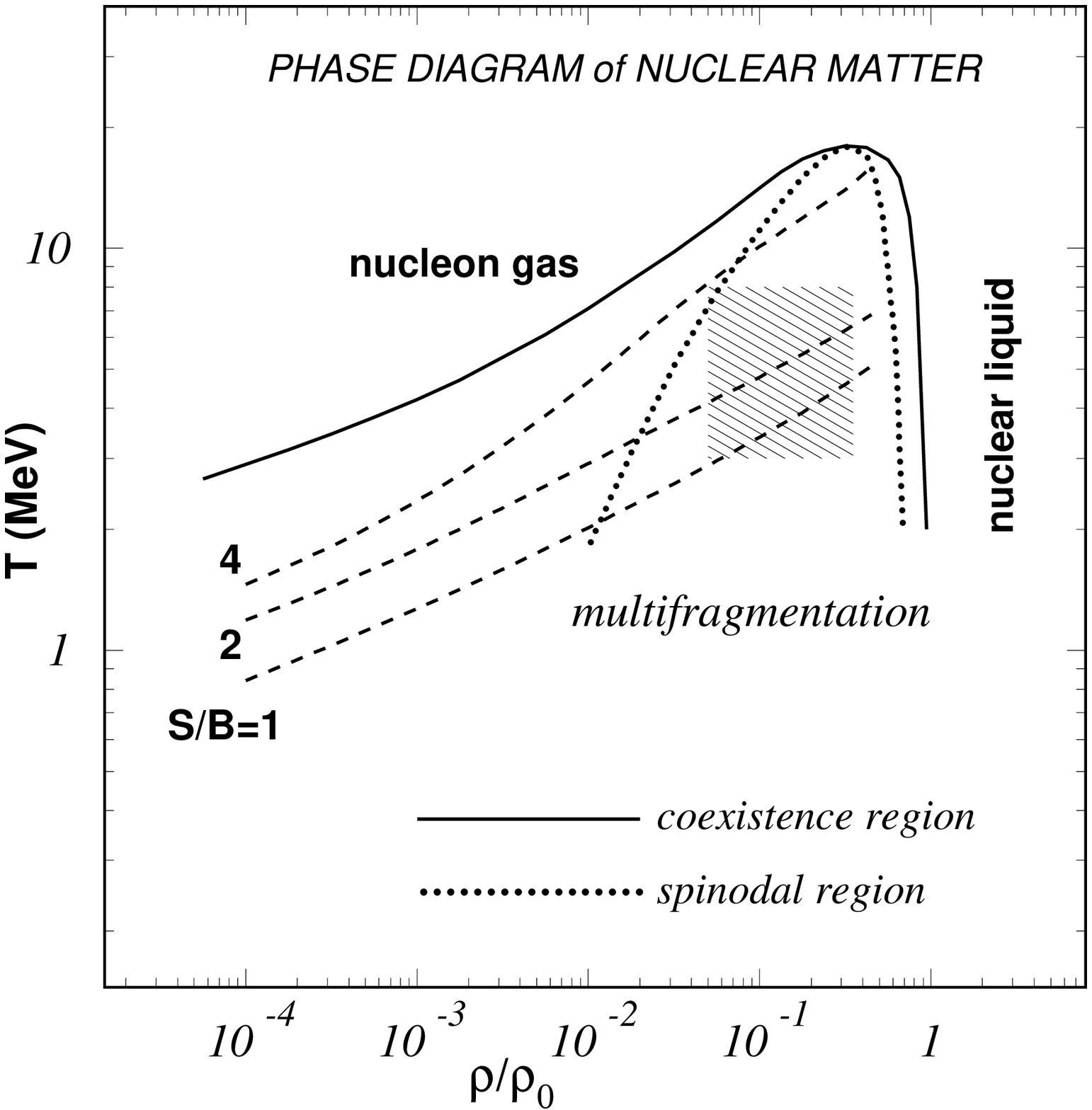}
\caption{\small{
Nuclear phase diagram on the 'temperature -- baryon density' 
plane. Solid and dotted lines indicate boundaries of the liquid-gas
coexistence region and the spinodal instability region. The shaded
area corresponds to typical conditions for nuclear multifragmentation
reactions. The dashed lines are isentropic trajectories characterized
by constant entropy per baryon, $S/B=$1, 2, and 4. 
}}
\end{figure} 

\begin{figure} [tbh]
\vspace{-1.3cm}
\hspace{2.4cm}
\includegraphics[width=8.6cm]{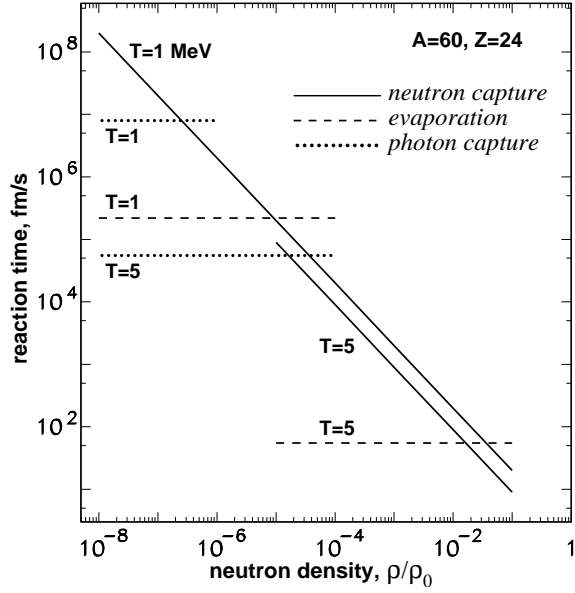}
\caption{\small{
Estimated reaction rates for a typical nucleus, with mass number $A$=60 
and charge $Z$=24, in stellar environment versus the density of free 
neutrons. 
Different lines correspond to different reaction types as indicated in the 
figure. 
Temperatures $T$ (in MeV) are given at the lines.
}}
\end{figure} 

\begin{figure} [tbh]
\vspace{-0.8cm}
\hspace{2cm}
\includegraphics[width=10cm]{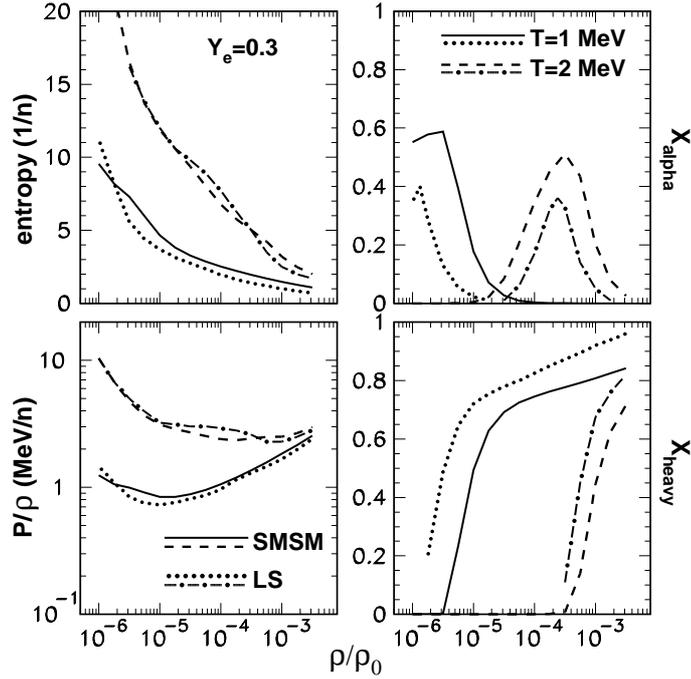}
\caption{\small{
Comparison of SMSM and the Lattimer-Swesty (LS) model \cite{Lattimer} for 
stellar matter with electron fraction $Y_e$=0.3 and temperatures 
$T=$1 and 2 MeV, as functions of the baryon density in units of the normal 
nuclear density $\rho_0\approx$0.15 fm$^-3$. The panels 
present the total entropy per nucleon, the pressure divided by density, 
the fractions of $\alpha$-particles and heavy nuclei ($A>4$). 
}}
\end{figure} 

\begin{figure} [tbh]
\vspace{-1cm}
\hspace{2cm}
\includegraphics[width=10cm]{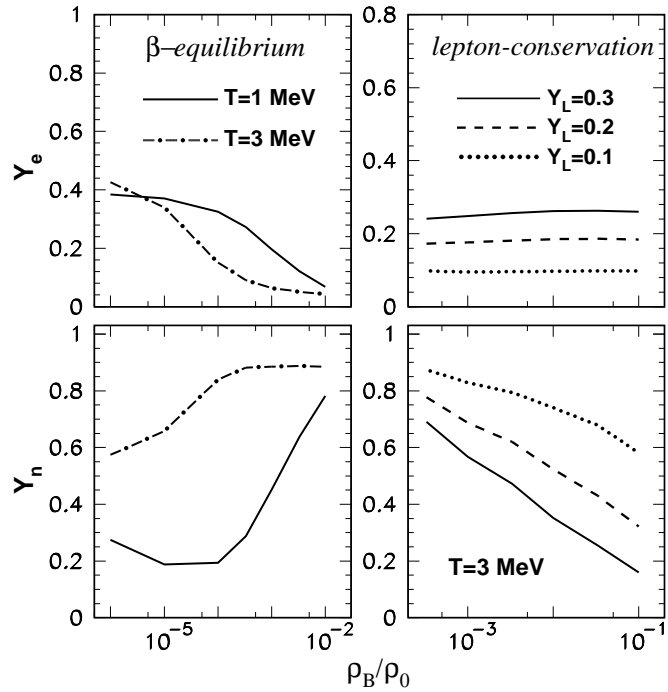}
\caption{\small{
Average fractions of electrons $Y_e$ (top panels) and
free neutrons $Y_n$ (bottom panels) versus baryon density in units of 
$\rho_0$. Left panels present results for the $\beta$-equilibrated matter 
with $T=$1 and 3 MeV. Right panels correspond to the conserved lepton 
fractions $Y_L$=0.1, 0.2 and 0.3, and $T=$3 MeV. 
}}
\end{figure} 

\begin{figure} [tbh]
\vspace{-1.2cm}
\hspace{2cm}
\includegraphics[width=10cm]{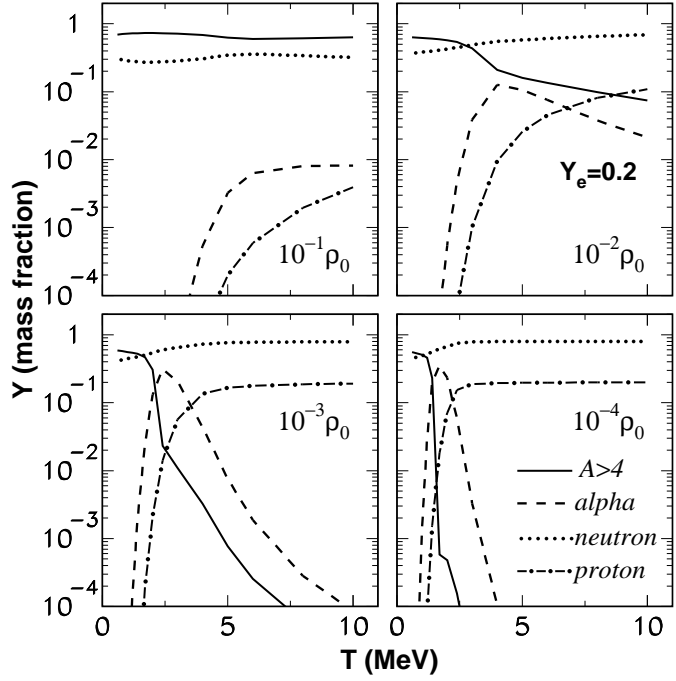}
\caption{\small{
Mass fractions of nuclear species for stellar matter at $Y_e$=0.2 as 
functions of temperature. 
Solid lines are for heavy nuclei ($A>4$), dashed lines -- 
$\alpha$-particles, dotted lines -- neutrons, dot-dashed lines -- protons. 
The results for baryon densities of $10^{-1}$, $10^{-2}$, 
$10^{-3}$, $10^{-4}\rho_0$ are presented in the corresponding panels. 
}}
\end{figure} 

\begin{figure} [tbh]
\vspace{-0.8cm}
\hspace{2cm}
\includegraphics[width=10cm]{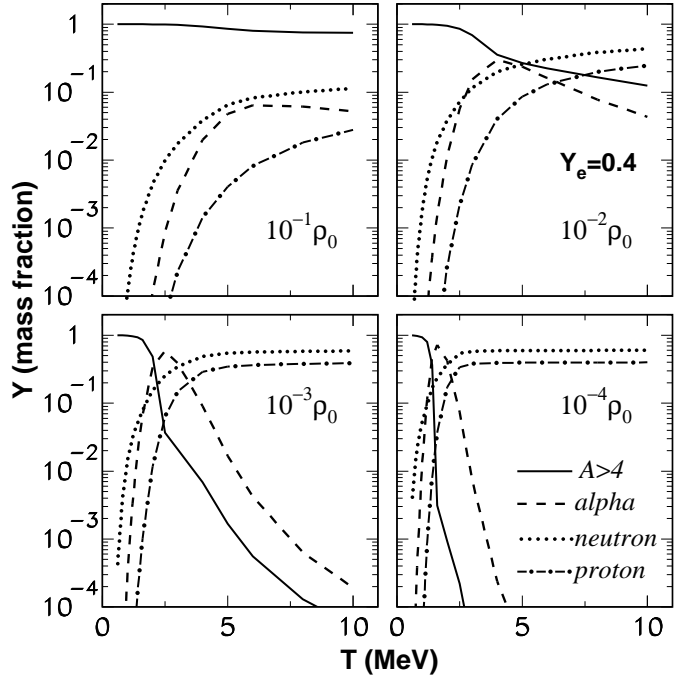}
\caption{\small{
The same as Fig.~5 but at $Y_e$=0.4. 
}}
\end{figure} 

\begin{figure} [tbh]
\vspace{-1.2cm}
\hspace{2cm}
\includegraphics[width=10cm]{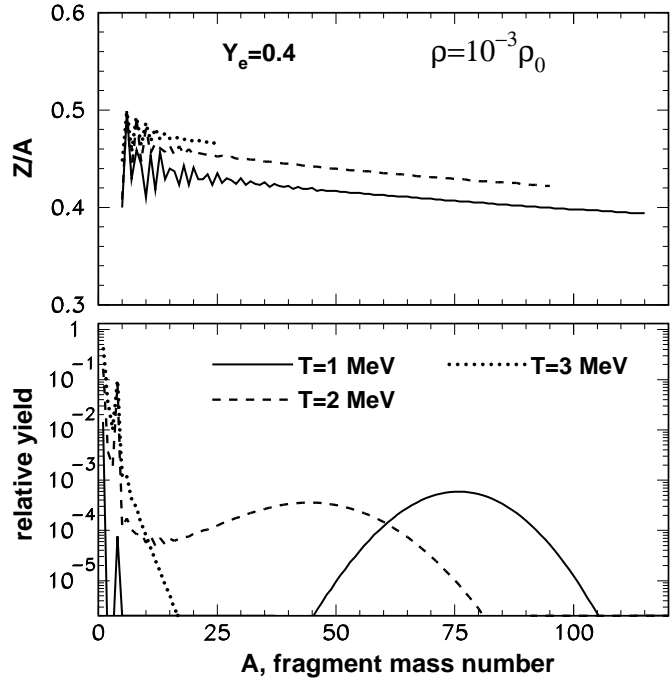}
\caption{\small{
Mass distributions, i.e., yields per nucleon, (bottom panel) and average 
charge to mass-number ratios (top panel) for nuclei produced at density 
$10^{-3}\rho_0$ and $Y_e$=0.4, for temperatures $T$=1, 2 and 3 MeV. 
}}
\end{figure} 

\begin{figure} [tbh]
\vspace{-0.8cm}
\hspace{2cm}
\includegraphics[width=10cm]{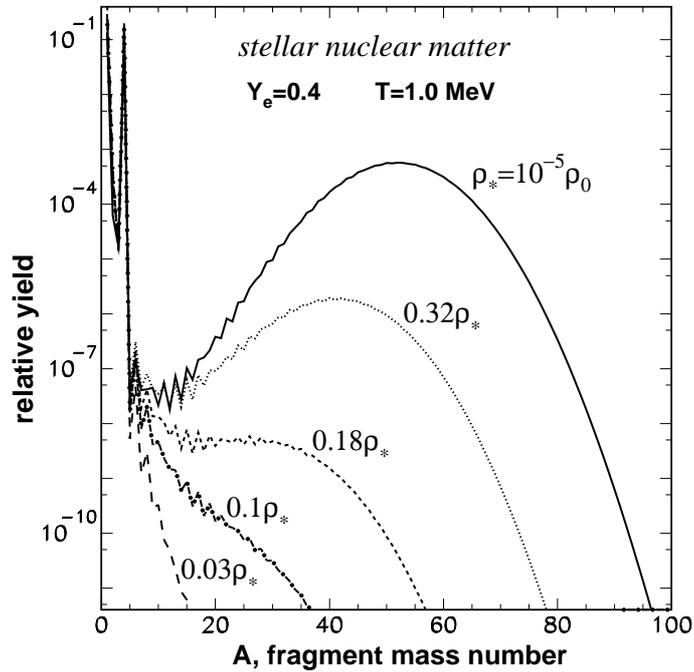}
\caption{\small{
Fragment mass distributions (yields per nucleon) at T=1 MeV and several 
densities in units of $\rho_{*}=10^{-5}\rho_0$ (see notations at the lines). 
Electron fraction is $Y_e$=0.4. 
}}
\end{figure} 

\begin{figure} [tbh]
\vspace{-1.5cm}
\hspace{2cm}
\includegraphics[width=10cm]{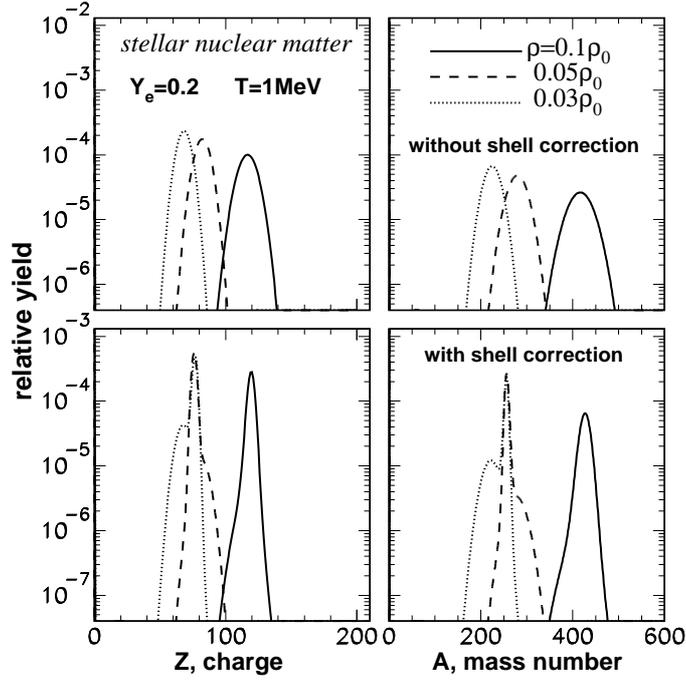}
\caption{\small{
Mass (right panels) and charge (left panels) distributions of 
superheavy nuclei at subnuclear densities without (top panels) 
and with (bottom panels) shell corrections. The stellar matter 
has electron fraction $Y_e$=0.2 and temperature $T$=1 MeV. 
Different lines correspond to different densities as indicated in 
the figure. 
}}
\end{figure} 

\begin{figure} [tbh]
\vspace{-0.5cm}
\hspace{2cm}
\includegraphics[width=10cm]{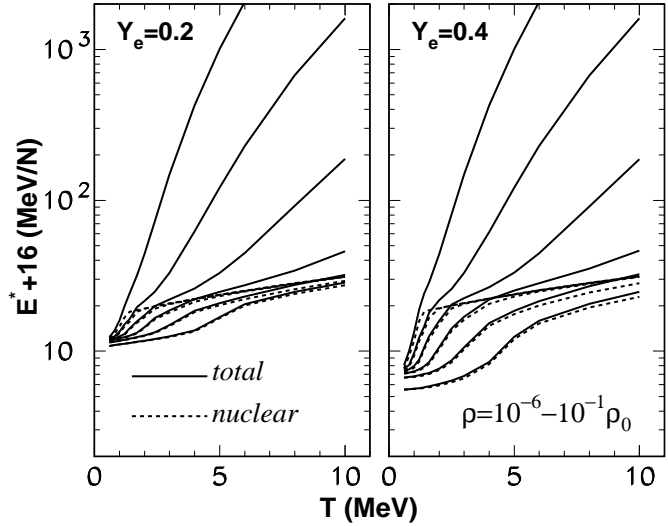}
\caption{\small{
Energy per nucleon as function of temperature in stellar 
environment with electron fractions $Y_e$=0.2 (left) and $Y_e$=0.4 (right), 
measured from the binding energy of normal nuclear matter 
(16 MeV per nucleon). Dashed 
lines show only contributions of nuclei. Solid lines give total energies 
including nuclear, electron and photon contributions. Baryon densities of 
10$^{-6}$, 10$^{-5}$, 10$^{-4}$, 10$^{-3}$, 10$^{-2}$, and 10$^{-1}$$\rho_0$, 
correspond to the 6 lines from the top to the bottom in the both panels. 
}}
\end{figure} 

\begin{figure} [tbh]
\vspace{-1.5cm}
\hspace{2cm}
\includegraphics[width=10cm]{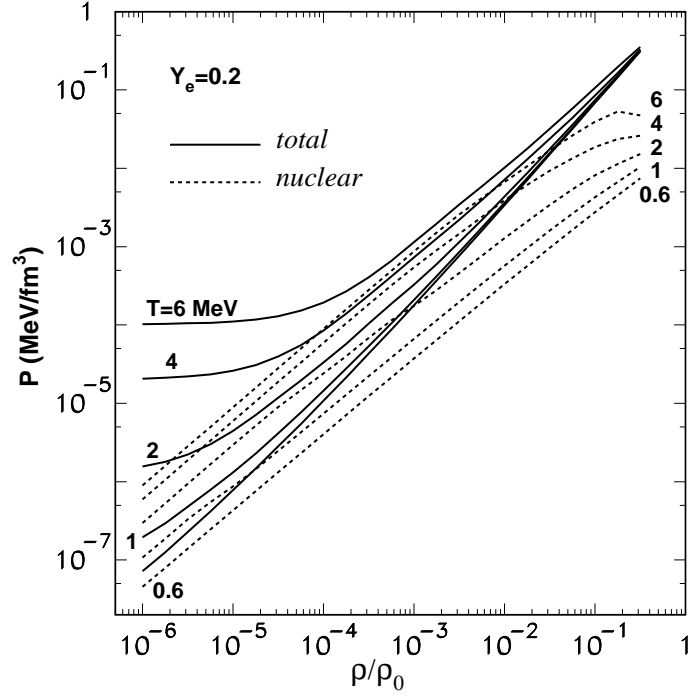}
\caption{\small{
Pressure as function of baryon density in units $\rho_0$ 
(the phase diagram in the $P-\rho$ plane) for $Y_e$=0.2. 
Dashed lines are only 
nuclear contributions, solid lines are total pressures including also 
electron and photon contributions. The temperatures of 6, 4, 2, 1, and 
0.6 MeV, are presented by different lines from the top to the bottom. 
}}
\end{figure} 

\begin{figure} [tbh]
\vspace{-0.5cm}
\hspace{2cm}
\includegraphics[width=10cm]{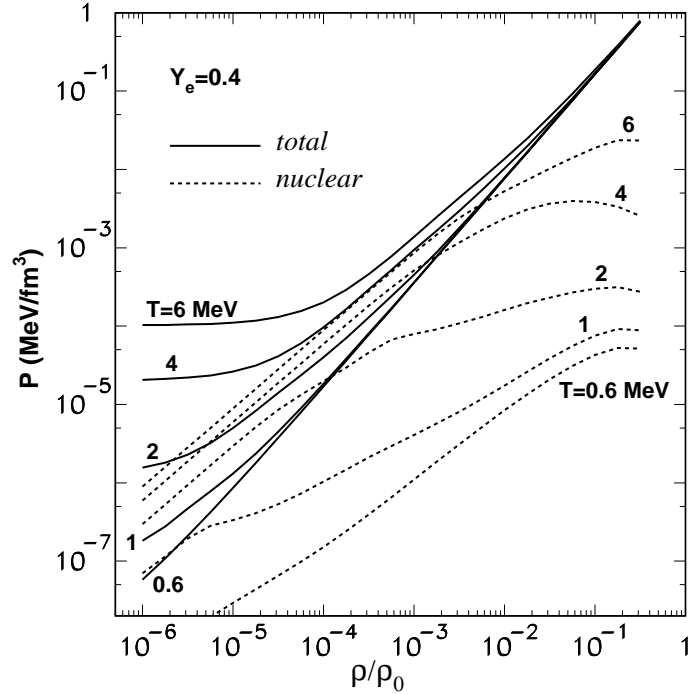}
\caption{\small{
The same as Fig.~11 but for $Y_e$=0.4. 
}}
\end{figure} 

\begin{figure} [tbh]
\vspace{-1cm}
\hspace{2cm}
\includegraphics[width=10cm]{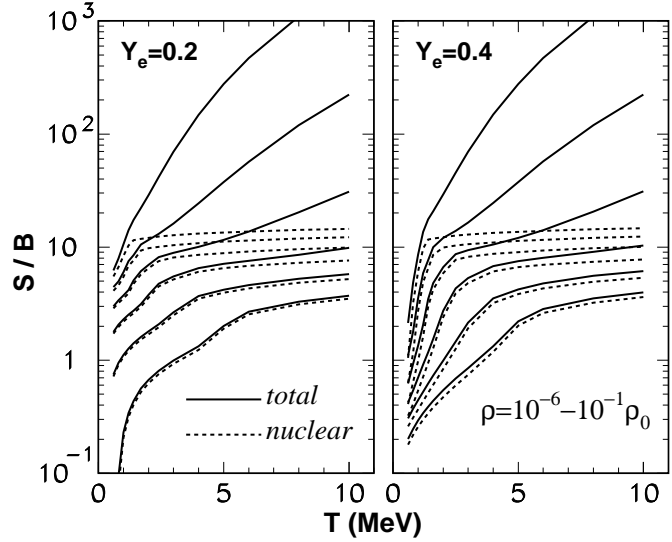}
\caption{\small{
Entropy (per baryon) of stellar matter as function of temperature, for 
electron fractions $Y_e$=0.2 (left) and $Y_e$=0.4 (right). Dashed lines 
show only nuclear contributions, solid lines give total entropy including 
also electron and photon contributions. The densities of 10$^{-6}$, 10$^{-5}$, 
10$^{-4}$, 10$^{-3}$, 10$^{-2}$, and 10$^{-1}$$\rho_0$, correspond to the 6 
lines from the top to the bottom in the both panels. 
}}
\end{figure} 

\begin{figure} [tbh]
\vspace{-1cm}
\hspace{2cm}
\includegraphics[width=10cm]{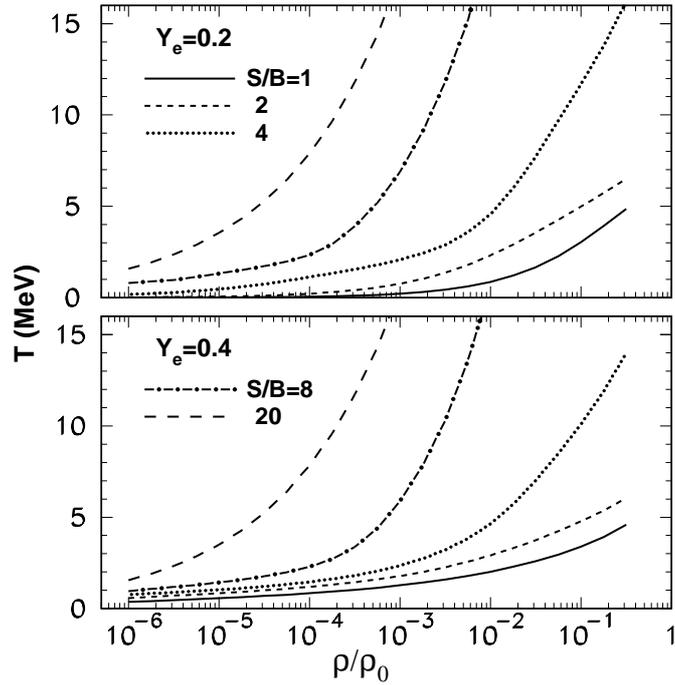}
\caption{\small{
Isentropic trajectories on the 'temperature -- baryon density' plane for 
different entropy per baryon values (S/B=1, 2, 4, 8, 20) indicated in the 
figure. Electron fractions are $Y_e$=0.2 (top) and $Y_e$=0.4 (bottom). 
}}
\end{figure} 

\clearpage

\begin{figure} [tbh]
\vspace{-1cm}
\hspace{2cm}
\includegraphics[width=10cm]{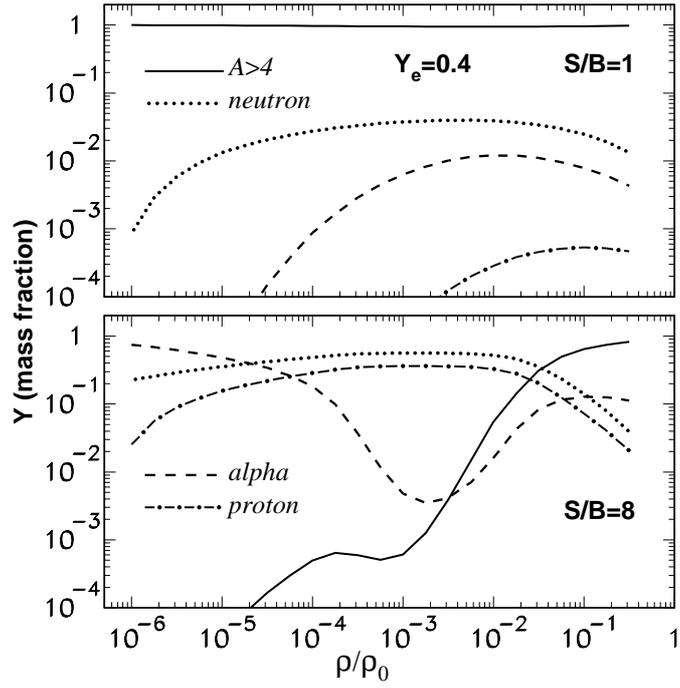}
\caption{\small{
Mass fractions of nuclear species along isentropes with S/B= 1 (top) and 
8 (bottom), for $Y_e$=0.4. Solid, dotted, dashed, and dot-dashed lines 
correspond to heavy nuclei ($A>4$), neutrons, $\alpha$-particles, and 
protons, respectively. 
}}
\end{figure} 

\begin{figure} [tbh]
\vspace{-1cm}
\hspace{2cm}
\includegraphics[width=10cm]{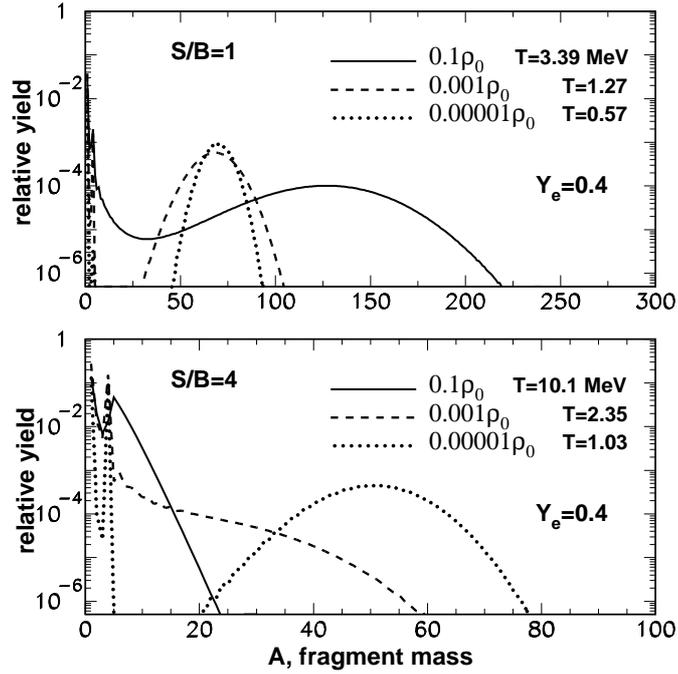}
\caption{\small{
Mass distributions of nuclei along isentropic trajectories with S/B= 1 (top) 
and 4 (bottom), for $Y_e$=0.4. 
Different lines correspond to specific baryon densities and temperatures 
(in MeV), as indicated in the figure. 
}}
\end{figure} 

\begin{figure} [tbh]
\vspace{-1cm}
\hspace{2cm}
\includegraphics[width=10cm]{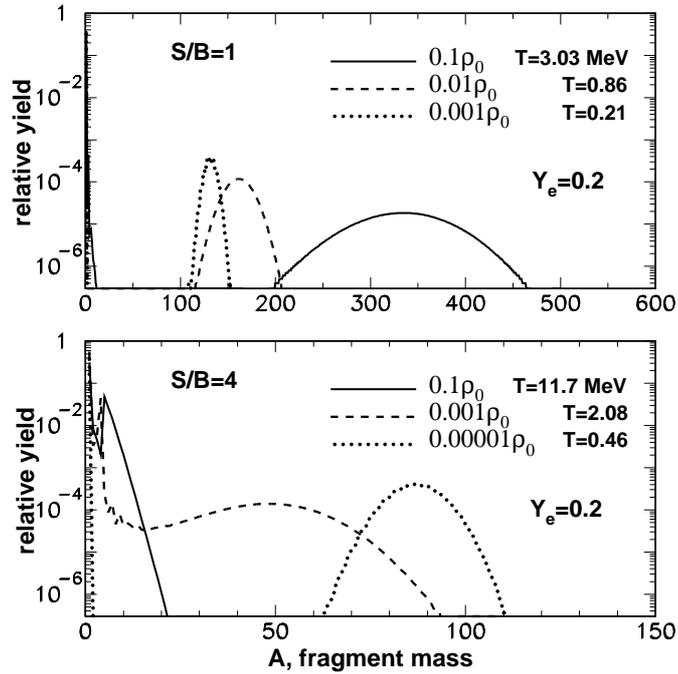}
\caption{\small{
The same as Fig.~16 but for $Y_e$=0.2. 
}}
\end{figure} 

\begin{figure} [tbh]
\vspace{-1.3cm}
\hspace{2cm}
\includegraphics[width=10cm]{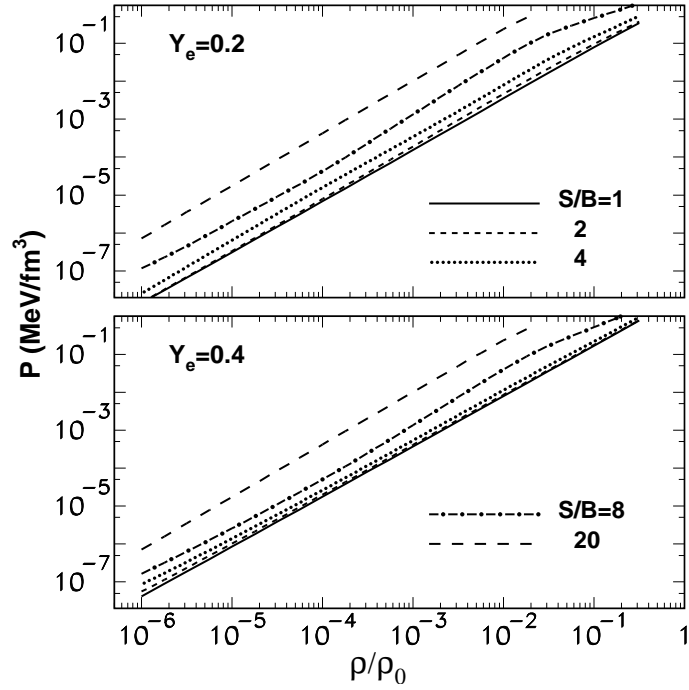}
\caption{\small{
Adiabatic pressure as function of baryon density (in units of $\rho_0$), 
for $Y_e$=0.2 (top) and $Y_e$=0.4 (bottom). Different lines correspond 
to entropies per baryon S/B=1, 2, 4, 8, and 20 units, as indicated in the 
figure. 
}}
\end{figure} 

\begin{figure} [tbh]
\vspace{-0.7cm}
\hspace{2cm}
\includegraphics[width=10cm]{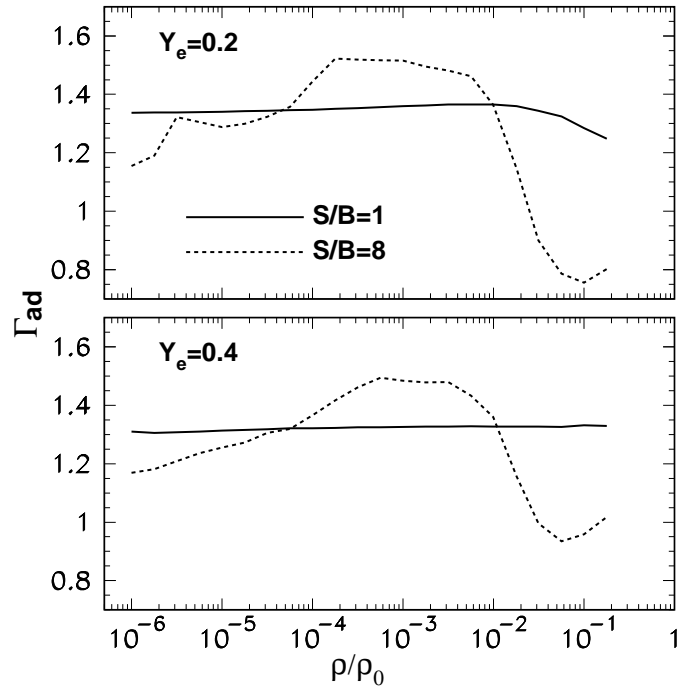}
\caption{\small{
Adiabatic index as function of baryon density at S/B= 1 (solid 
lines) and S/B=8 (dashed lines), for $Y_e$=0.2 (top) and  $Y_e$=0.4 (bottom).
}}
\end{figure} 

\begin{figure} [tbh]
\vspace{-1cm}
\hspace{2cm}
\includegraphics[width=10cm]{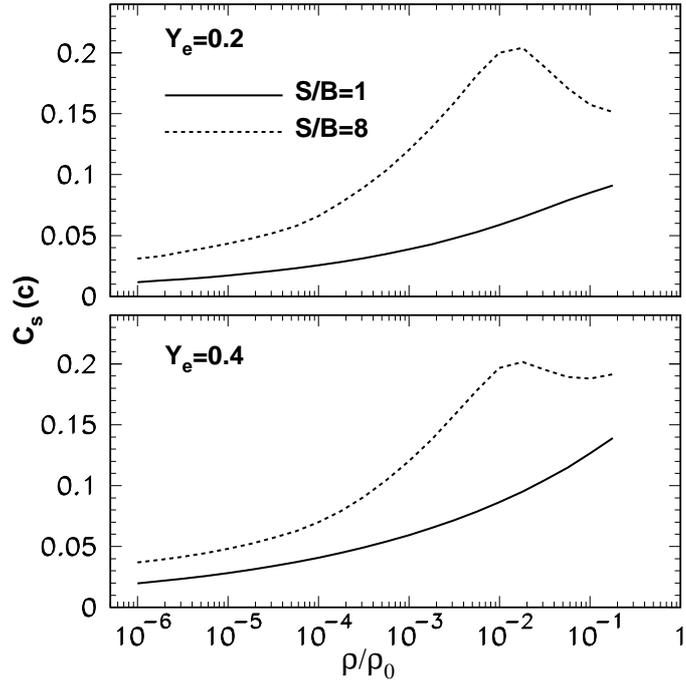}
\caption{\small{
Adiabatic sound velocity $c_s$ (in units of light velocity) at 
S/B= 1 (solid lines) and S/B=8 (dashed lines), for $Y_e$=0.2 (top) 
and $Y_e$=0.4 (bottom). 
}}
\end{figure} 

\begin{figure} [tbh]
\vspace{-1cm}
\hspace{2.3cm}
\includegraphics[width=10cm]{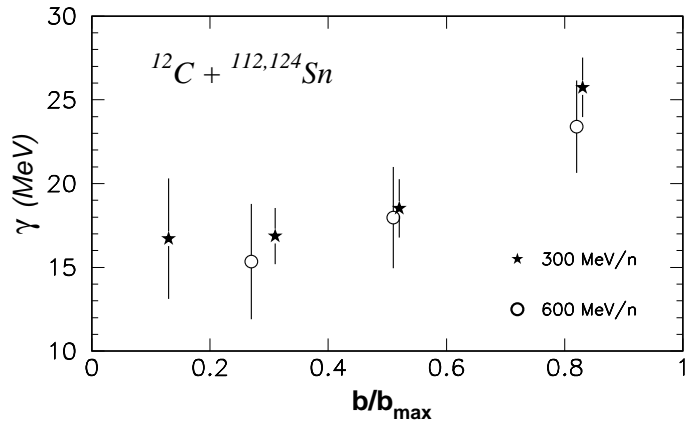}
\caption{\small{
The apparent symmetry energy coefficient
$\gamma$ of hot nuclei, as extracted from multifragmentation of tin
isotopes induced by $^{12}C$ beams with energy 300 and 600
MeV per nucleon, versus relative impact parameter $b/b_{max}$
\cite{LeFevre}.
}}
\end{figure} 

\begin{figure} [tbh]
\vspace{-1.5cm}
\hspace{2cm}
\includegraphics[width=10cm]{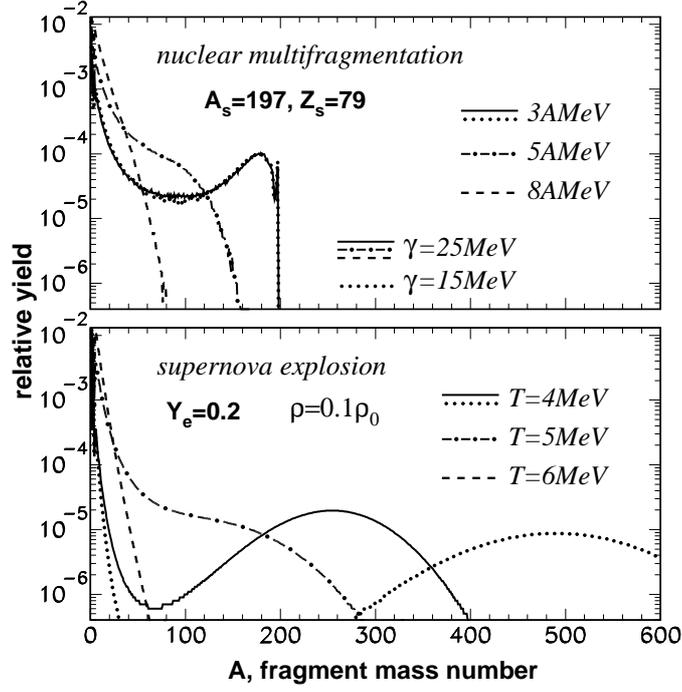}
\caption{\small{
Fragment mass distributions (yields per nucleon) in
multifragmentation of $Au$ sources (top panel) and
in supernova environment at the electron fraction $Y_e=$0.2 and 
baryon density 0.1$\rho_0$ (bottom panel). The 
calculations at excitation energies of 3, 5, and 8 MeV per nucleon (top),
and different temperatures $T$ (bottom), are shown by different curves.
Effects of the reduced symmetry energy coefficient $\gamma$ are also 
demonstrated in both panels. 
}}
\end{figure} 

\begin{figure} [tbh]
\vspace{-0.4cm}
\hspace{2.5cm}
\includegraphics[width=8.8cm]{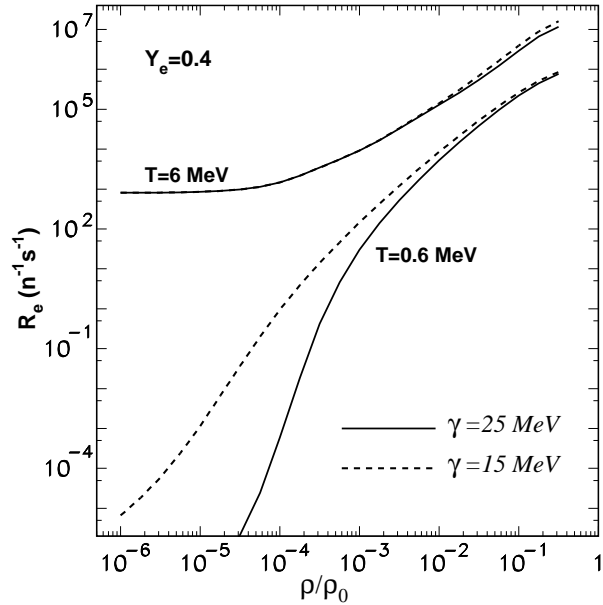}
\caption{\small{
Density dependence of electron-capture rates $R_e$
on hot nuclei in supernova environment at different
temperatures $T$ and the electron fraction $Y_e$=0.4.
Solid and dashed lines show results for standard (25 MeV) and
reduced (15 MeV) values of symmetry energy coefficients $\gamma$.
}}
\end{figure} 

\end{document}